\documentstyle[11pt]{article}
\normalbaselines
\begin{document}
\bibliographystyle{unsrt}

\def\bea*{\begin{eqnarray*}}
\def\eea*{\end{eqnarray*}}
\def\ba{\begin{array}}
\def\ea{\end{array}}
\count1=1
\def\be{\ifnum \count1=0 $$ \else \begin{equation}\fi}
\def\ee{\ifnum\count1=0 $$ \else \end{equation}\fi}
\def\ele(#1){\ifnum\count1=0 \eqno({\bf #1}) $$ \else \label{#1}\end{equation}\fi}
\def\req(#1){\ifnum\count1=0 {\bf #1}\else \ref{#1}\fi}
\def\bea(#1){\ifnum \count1=0   $$ \begin{array}{#1}
\else \begin{equation} \begin{array}{#1} \fi}
\def\eea{\ifnum \count1=0 \end{array} $$
\else  \end{array}\end{equation}\fi}
\def\elea(#1){\ifnum \count1=0 \end{array}\label{#1}\eqno({\bf #1}) $$
\else\end{array}\label{#1}\end{equation}\fi}
\def\cit(#1){
\ifnum\count1=0 {\bf #1} \cite{#1} \else 
\cite{#1}\fi}
\def\bibit(#1){\ifnum\count1=0 \bibitem{#1} [#1    ] \else \bibitem{#1}\fi}
\def\ds{\displaystyle}
\def\hb{\hfill\break}
\def\comment#1{\hb {***** {\em #1} *****}\hb }

\newtheorem{theorem}{Theorem}
\newtheorem{lemma}{Lemma}
\newtheorem{proposition}{Proposition}

\newcommand{\TZ}{\hbox{\bf T}}
\newcommand{\MZ}{\hbox{\bf M}}
\newcommand{\ZZ}{\hbox{\bf Z}}
\newcommand{\NZ}{\hbox{\bf N}}
\newcommand{\RZ}{\hbox{\bf R}}
\newcommand{\CZ}{\,\hbox{\bf C}}
\newcommand{\PZ}{\hbox{\bf P}}
\newcommand{\QZ}{\hbox{\bf Q}}
\newcommand{\HZ}{\hbox{\bf H}}
\newcommand{\EZ}{\hbox{\bf E}}
\newcommand{\GZ}{\,\hbox{\bf G}}

\vbox{\vspace{38mm}}
\begin{center}
{\LARGE \bf Baxter's T-Q Relation and Bethe Ansatz
of \\ [3mm] Discrete Quantum Pendulum and
Sine-Gordon Model  }\\[10 mm]
Shao-shiung Lin
\\ {\it Department of Mathematics \\ 
Taiwan University \\ Taipei, Taiwan \\
( email: lin@math.ntu.edu.tw ) }\\[3 mm]
Shi-shyr Roan \\
{\it Institute of Mathematics \\
Academia Sinica \\  Taipei , Taiwan \\
(email: maroan@ccvax.sinica.edu.tw ) } \\[30mm]
\end{center}

\begin{abstract}
Using the Baxter's $T$-$Q$ relation derived from
the transfer matrix technique, we
consider the diagonalization problem of  
discrete quantum pendulum and discrete quantum
sine-Gordon Hamiltonian from the algebraic
geometry aspect. For a finite chain system of
the size $L$, when the spectral curve 
degenerates into rational curves, we have reduced
Baxter's $T$-$Q$ relation into a polynomial
equation; the
 connection of $T$-$Q$ polynomial equation with
the algebraic Bethe Ansatz is clearly established
. In particular, for $L=4$ it is the case of
rational spectral curves for the discrete quantum
pendulum  and discrete sine-Gordon model. To these
Baxter's
$T$-$Q$ polynomial equations,  we have
obtained the complete and explicit solutions 
with a detailed understanding of their
quantitative and qualitative
 structure. In general the model possesses a 
spectral curve with a generic parameter. We have
conducted certain qualitative study on
the algebraic geometry of this high-genus Riemann
surface incorporating  Baxter's $T$-$Q$ relation. 
\end{abstract}

\vfill
\eject

\section{Introduction}
In the early seventies, R. Baxter proposed
the method of $Q$-operator and the $T$-$Q$
relation in his renowned solution of the
eight-vertex  model and the 
spin $\frac{1}{2}$ XYZ chain in soluble
statistical mechanics  \cite{B, Bax}. Since then,
the method has played a powerful mechanism up to
nowadays in the two-dimensional exactly solvable
lattice models, and the corresponding 
quantum spin-chain Hamiltonians. 
The quantum inverse scattering /algebraic Bethe
Ansatz method developed by the  Leningrad
school in the early eighties \cite{F, KS} 
systematized
earlier results on two-dimensional integrable 
lattice models, and paved the way for the
far-reaching effects in both mathematical and
physical development in the past two decades.
In the study of massless lattice
sine-Gordon model, Izergin-Korepin and Tarasov 
\cite{IK} found, within the framework of quantum
inverse scattering method, the
${\cal L}$-operator ( with $\CZ^N$-operators
entries) which satisfies the
Yang-Baxter equation for the six-vertex
model's $R$-matrix 
 with the parameter $q$ being the
$N$th root of unity:
$ q^N=1$, ( see
the formula (\req(LUV)) of this
paper). A slightly modified version of this ${\cal
L}$-operator also appeared in the study of 
chiral Potts
$N$-state model 
\cite{BBP}. On the other hand, there are 
Hamiltonians of physical interest,
exhibiting an intimate relationship with certain
systems derived from the transfer matrix for a
fixed finite size $L$ while
$N$ varying. For
$L$=3, the Hamiltonian, proposed first in 
\cite{FK} and then investigated in a
rigorously mathematical manner in \cite{LR},
can be interpreted as a generalization of  the
Hofstadter model
\cite{A, H}, a renowned Bloch system with a
constant external magnetic field. Through the
quantum inverse scattering method, one can
calculate the spectrum of the Hamiltonian by
solving the (algebraic) Bethe ansatz equation. In
\cite{LR} we formulate the problem from the
algebraic geometry aspect, and consider it as the
Baxter's
$T$-$Q$ relation on
the spectral curve via the Baxter's vacuum state
\cite{Bax, BazS}.         
In addition, 
a general scheme of diagonalizing the transfer
matrix of a finite size $L$ 
 by means of the Baxter's
$T$-$Q$ relation ( or the Bethe equation) on
the spectral curve was also discovered.
At present, it is rather difficult to
extract the explicit quantitative data for
the spectrum problem by this approach, due to the
complicated functional theory of the high-genus
spectral curve. Nevertheless,  when the
spectral curve is totally degenerated into
rational curves, we demonstrate in this paper
that the polynomial equation derived from the 
Baxter's
$T$-$Q$ relation gives rise  to the usual Bethe
ansatz equation in physical literature. And for
$L=4$,  the transfer matrix produces the
Hamiltonions of the discrete quantum pendulum and
the discrete quantum sine-Gordon (SG) model,
appeared in 
\cite{BKP, KKS}.   

In this article we make a thorough study of the
discrete quantum pendulum and SG model in the
context of Baxter's
$T$-$Q$ relation by using the transfer
matrix technique, ( for the Hamiltonians, see
(\req(DQP)) (\req(SG)) of Sect. 2). Though 
the formulation via this approach can be made on
Hamiltonian chains of an arbitrary  finite
size $L$, the mathematical treatment given
in this paper for these two specific models takes
advantage of certain special features  presented
only in the size $L$=4. The spectral curves, upon
which the Baxter's
$T$-$Q$ relation is formulated, have a very high
genus in general; indeed it is of the order 
$N^5$. However, by examining
their algebraic geometry properties, these
spectral curves are found to form a family of
algebraic curves covering the elliptic curves, so
one could hope that the elliptic function theory
would eventually play a role in the solutions of 
Baxter's
$T$-$Q$ relation. In the rational-spectral-curve
case for a finite size $L$, the
 geometry of spectral curves becomes trivial.
However, it still inevitably requires some subtle
analysis of the Baxter vacuum state to extract the
essential data on 
solutions of the polynomial  equation, then
to carry out the necessary algebraic study of a
certain "over-determined" system of 
$q$-difference  equations for a $N$th root of
unity $q$, a difficult problem  
for a general $L$. In the cases of 
discrete quantum pendulum and 
SG model, where the size $L$ is equal to 4,  
we are able to obtain the explicit  solutions of
Baxter's $T$-$Q$ polynomial equation by taking
 the special symmetric  structure of
polynomial functions into account. The
results are  complete from both the quantitative
and qualitative aspects, and the method provides a
sound mathematical treatment on problems
raised in
\cite{BKP, KKS}. The main advantage of our
approach to these problems is to make use of the
explicit  Baxter's vacuum state as the precise
form of "Bethe ansatz ground state", which was
only postulated in the previous articles in
literature, e.g. \cite{KKS}. This approach 
to the eigenvalue problem of Hamiltonians
has also  been shown in
\cite{LR} for $L$=3 on the Hofstadter-type models.
It seems to us that  the use of Baxter's
$T$-$Q$ relation would be more
fundamental, and mathematically tractable, 
than the usual Bethe Ansatz technique.
Furthermore, in this approach a certain
mathematical theory of
$q$-Sturm-Liouville type problem would possibly
arise to entangle with the Baxter's
$T$-$Q$ polynomial equation originated from
certain physical problems.

This paper is organized as
follows:  In Sect. 2, we review the basic
construction of Baxter's
$T$-$Q$ relation using the transfer matrix 
techniques, with the spectral curve of a high
genus  depending on the size $L$ and 
parameters involved in the Hamiltonions.  Some
facts in
\cite{LR} will be recalled  here for the sake of
completeness. Then we describe the constraints of
parameters appeared in the spectral curves for the
discrete quantum pendulum and SG model. In
Sect. 3, we discuss a canonical 
 procedure of reducing the Baxter's
$T$-$Q$ relation to a polynomial equation  when
the spectral curve is degenerated to  rational
curves for a finite size $L$. By
converting the parameters to one special case
studied in our earlier paper
\cite{LR}, we obtain the Baxter's $T$-$Q$
 polynomial equation. In Sect. 4,  we
apply the conclusion of the previous section to
the case $L$=4, and with the further parameter
constraints of the discrete  quantum pendulum and
SG model. The symmetric Baxter's $T$-$Q$
polynomial relation is introduced, and a general
discussion on the qualitative nature of its
solutions is given.  In Sect. 5, we 
construct explicitly the complete solutions of 
symmetric Baxter's
$T$-$Q$ polynomial equations, among which the
discrete  quantum pendulum and SG
model for the
rational spectral curve are included. Both
the quantitative and qualitative nature of these
solutions are revealed in the procedure of the
derivation; in particular, the
connection of the solutions with the usual
Bethe ansatz  equation in literature is
clarified.  In Sect. 6, we consider the  discrete
quantum pendulum and SG model for the 
general spectral curves, which are high-genus
Riemann surfaces. By examining the geometric
properties, we discover a canonical
relationship of these curves with elliptic
curves. A primitive qualitative analysis is made
on the geometry of these spectral curves in
connection with the eigenvalue problem of 
physical models through  Baxter's
$T$-$Q$ relation.  In Sect. 7, we present the
conclusion remark with a discussion of our future
programs.
We end with the appendix of presenting the
identification of the
sine-Gordon integral in
\cite{KKS} with the Hamiltonion given in this
paper.

{\bf Notations. } To present 
our work, we prepare some notations. In this
paper, 
$\ZZ, \RZ, \CZ$ will denote 
the ring of integers, real, complex numbers
respectively, $\NZ = \ZZ_{>0}$, $\ZZ_N=
\ZZ/N\ZZ$,  and ${\rm i} = \sqrt{-1}$. For a positive integers $n$, we denote
$\stackrel{n}{\otimes} \CZ^N$ the tensor
product of $n$-copies of the vector space $\CZ^N$.
We use the notation of ${\sf q}$-shifted 
factorials,
$$
(a ; {\sf q} )_0 = 1, \ \ \ \
(a ; {\sf q} )_n = 
(1-a)(1-a {\sf q}) \cdots (1-a
{\sf q}^{n-1}) , \ \ \ \ n \in \NZ  \ .
$$
We shall minimize the repetition of materials
in our previous article \cite{LR}, and so opt
to use the same notation conventions as much as
possible.

\section{Transfer Matrix, Baxter Vacuum State
and T-Q Equation}
In this section we first recall some formulae
in quantum inverse scattering method for later
use. Most of the materials can be found in
\cite{LR}, including the original references.
Then, we specify our discussion on
the cases of the discrete quantum
pendulum and SG Hamiltonian, the
models we shall mainly consider in this work.

In this paper, $N$ will always denote an odd
positive  integer with $M= [\frac{N}{2}]$: 
$N= 2M+1 \ ( M \geq 1)$,  
and $\omega$ is a primitive
$N$th root of unity,  $q:=
\omega^{\frac{1}{2}}$ with   
$q^N = 1$, i.e.  
$q = \omega^{M+1 }$. A vector
$v $ in $\CZ^N$ is represented
by a sequence of coordinates, $(v_k \ | \  k
\in \ZZ)$, with the $N$-periodic condition: $v_k =
v_{k+N}$, i.e. $v = (v_k)_{k
\in \ZZ_N}$.  
The standard basis of $\CZ^N$ will be denoted by 
$|k\rangle \ (k \in \ZZ_N ) $, with the dual basis
$\langle k|  \ (k \in \ZZ_N )$  of $\CZ^{N*}$.

Let $Z, X$ be a pair of generators of the Weyl
algebra with  the
$N$th power identity : $
ZX= \omega XZ, \ 
Z^N = X^N = I$,  
and denote by $Y (: =ZX )$ the element in the Weyl
algebra. One has the relations,  $
XY = \omega^{-1} YX$, $  YZ = \omega^{-1} ZY$,
and $Y^N= 1$. 
The canonical  
representation of the Weyl algebra is the
(unique) $N$-dimensional irreducible
representation, with the following matrix
realization:
$$
Z(v)_k = \omega^n v_k \ ,
\ \  X(v)_k = v_{k-1} \ , \
\ \ Y(v)_k = \omega^k
v_{k-1} \ \ \ {\rm for} \ \ v=(v_k) \in \CZ^N .
$$
By using the operators $X, Y, Z$ and the identity
$I$, one can form   a solution of the Yang-Baxter
(YB) equation for a  slightly modified 
$R$-matrix of the  six-vertex model, appeared
first in \cite{FK}, and then 
studied in great details in \cite{LR}.
The 
$L$-operator, depending on a parameter
$h=[a:b:c:d]$ in the projective 3-space $ \PZ^3$,
is given by the following
$2 \times 2$ matrix with operator-valued entries
acting on the quantum space
$\CZ^N$:    
\be
L_h (x) = \left( \begin{array}{cc}
       aY  & xbX  \\
        xcZ &d    
\end{array} \right) \ , \ \ x \in \CZ \ ,
\ele(L)
and it satisfies the following YB
relation:
\begin{eqnarray}
R(x/x') (L_h (x) \bigotimes_{aux}1) ( 1
\bigotimes_{aux} L_h (x')) = (1
\bigotimes_{aux} L_h (x'))(L_h (x)
\bigotimes_{aux} 1) R(x/x') \ ,
\label{eq:RLL}
\end{eqnarray}
where the script letter $"aux"$ indicates an 
operation taking on the auxiliary space $\CZ^2$,  
and $R(x)$ is the matrix of 2-tensor of the
auxiliary  space  with the following  numerical
expression,
$$
R(x) = \left( \begin{array}{cccc}
        x\omega-x^{-1}  & 0 & 0 & 0 \\
        0 &\omega(x-x^{-1}) & \omega-1 &  0 \\ 
        0 & \omega-1  &x-x^{-1} & 0 \\
     0 & 0 &0 & x\omega-x^{-1}     
\end{array} \right) \ .
$$
The operator (\req(L)) is related to the
the following $L$-operator appeared in
\cite{IK} for the study of  sine-Gordon lattice
model by using the  
six-vertex model ${\sf R}$-matrix  and  the Weyl
operators $U, V$ (
$UV = q^{-1} VU$,
$ U^N=V^N =1$) : 
\be
L^{\ast}_h (x) = \left( \begin{array}{cc}
       aqU  & xbV^{-1}  \\
        xcV &dU^{-1}    
\end{array} \right) \ , \ \ 
{\sf R}(x) = \left( \begin{array}{cccc}
        x q -x^{-1}q^{-1}  & 0 & 0 & 0 \\
        0 & x-x^{-1} & q-q^{-1} &  0 \\ 
        0 & q-q^{-1}  & x-x^{-1} & 0 \\
     0 & 0 &0 &  x q -x^{-1}q^{-1}    
\end{array} \right) 
\ele(LUV)
with the YB relation: $
{\sf R}(x/x')((L^{\ast}_h
(x)\bigotimes 1) \otimes (1 \bigotimes
L^{\ast}_h (x')) =  ((1 \bigotimes
L^{\ast}_h (x') \otimes (L^{\ast}_h (x)
\bigotimes 1))
{\sf R}(x/x')$. 
Indeed, by the identification: $
Z= VU$, $ X= V^{-1}U$,  ( equivalently, $ U =
q^{\frac{-1}{2}}Y^{\frac{1}{2}} ,  V =
q^{\frac{1}{2}} Z Y^{\frac{-1}{2}} $,) 
$L_h(x)$ in (\req(L)) and 
$L^{\ast}_h(x)$ in (\req(LUV)) are related by 
\be
L^{\ast}_h (x) =  L_h (x)Y^{\frac{-1}{2}}
q^{\frac{1}{2}}  \  . 
\ele(LsL)
By the matrix-product on 
auxiliary  spaces and tensor-product of 
quantum spaces, the following 
$L$-operator of a finite
size $L$, depending on the
parameter $\vec{h}= (h_0,
\ldots, h_{L-1}) \in  (\PZ^3)^L $,
\begin{eqnarray*}
L_{\vec{h}}(x) =
\bigotimes_{j=0}^{L-1}L_{h_j}(x) \ \ \ ( :=
L_{h_0}(x)
\otimes L_{h_1}(x) 
\otimes \ldots \otimes L_{ h_{L-1}}(x) ) \ ,
\end{eqnarray*}
again satisfies the YB relation 
(\ref{eq:RLL}), hence its trace gives rise to a
commuting family of transfer matrices: 
$$
T_{\vec{h}} (x) = {\rm tr}_{aux} ( L_{\vec{h}}(x))
\ , \ \ \ x \in \CZ \ .
$$
The same conclusion also holds for
$L^{\ast}_{\vec{h}} (x)$, and $ 
T^{\ast}_{\vec{h}} (x)$, defined by 
$$
L^{\ast}_{\vec{h}} (x) = 
\bigotimes_{j=0}^{L-1} L^{\ast}_{h_j} (x) \ , \ \
\ \   T^{\ast}_{\vec{h}} (x) ={\rm tr}_{aux}
(L^{\ast}_{\vec{h}} (x)) \ .
$$
By (\req(LsL)), one has the following relation of
the above two types of transfer matrices :
\be
T^{\ast}_{\vec{h}} (x)   = T_{\vec{h}}
(x)D^{\frac{-1}{2}} \ , \ \ 
{\rm where } \ \ \ D: = q^{-L}
\stackrel{L}{\otimes} Y .
\ele(TT*r)

For later use, we now summarize some basic facts
on the Baxter's $T$-$Q$ relation and the Baxter
vacuum state\footnote{The
Baxter's $T$-$Q$ relation and the Baxter
vacuum state here were  called by the Bethe
equation, the Baxter
vector respectively  in
\cite{FK, LR}} over the spectral curve in the
diagonalization problem of transfer matrix
$T_{\vec{h}}(x)$ (for the details, see
\cite{LR}). Applying the following
gauge transform  on 
$L_{h_j}(x) $:
$$
\widetilde{L}_{h_j}(x, \xi_j, \xi_{j+1}) = A_j
L_{h_j}(x) A_{j+1}^{-1}
\ ,\ \ {\rm where } \ \  A_j = \left(
\begin{array}{lc} 1 &  \xi_j-1\\
1 &  \xi_j 
\end{array}\right)  \ \  ( 0 \leq j \leq L-1 ) , \
\ {\rm and} \ \ A_{L}: = A_0 \ , 
$$
one has the expression:  
$$
\widetilde{L}_{h_j} (x, \xi_j, \xi_{j+1})  = \left(
\begin{array}{ll} F_{h_j}(x,  \xi_j -1, \xi_{j+1}) &
-F_{h_j} (x, 
\xi_j -1, \xi_{j+1} -1) \\ F_{h_j}(x,  \xi_j ,
\xi_{j+1}) & - F_{h_j}(x, 
\xi_j ,
\xi_{j+1}-1 )
\end{array}\right) \ .
$$
Here the operator $F_h (x,  \xi, \xi' )$ is
defined by 
$F_h (x,  \xi, \xi' ) :=
\xi'aY - xbX +  \xi'\xi xc Z -\xi d $.
Accordingly, for $\vec{h} \in (\PZ^3)^L$ and  
$\vec{\xi}= (\xi_0,  \ldots , \xi_{L-1} ) \in
(\CZ^N)^L$, the modified
$L$-operator becomes
\begin{eqnarray*}
\widetilde{L}_{\vec{h}}(x,
\vec{\xi} ) : = 
\bigotimes_{j=0}^{L-1}
\widetilde{L}_{h_j}(x,
\xi_j, \xi_{j+1})   = 
\left( \begin{array}{cc}
   \widetilde{L}_{ \vec{h}; 11}(x, \vec{\xi}) &
\widetilde{L}_{
\vec{h};12}(x, \vec{\xi})  \\
     \widetilde{L}_{ \vec{h}; 21}(x, \vec{\xi}) &
\widetilde{L}_{ \vec{h} ; 22}(x, \vec{\xi}) 
\end{array} \right) , \ {\rm and} \ \ \xi_L :=
\xi_0 
\ . 
\end{eqnarray*}
As the gauge-transform procedure keeps 
the trace unchanged, we have the relation:
$ T_{\vec{h}} (x) = {\rm tr}_{aux}
(\widetilde{L}_{\vec{h}}(x,
\vec{\xi} ) ) $. For a given $\vec{h}$, we will 
consider the variable $(x, \vec{\xi}) $  in
the curve
${\cal C}_{\vec{h}}$ defined by  
\begin{eqnarray}
{\cal C}_{\vec{h}} : \ \ \xi_j^N  =(-1)^N
\frac{\xi_{j+1}^Na_j^N - x^Nb_j^N}{\xi_{j+1}^N x^N
c_j^N - d_j^N } \ \ , \ \ \ \  j =0,
\ldots, L-1 \ . \label{eq:Cvh}
\end{eqnarray} 
We shall call ${\cal C}_{\vec{h}}$ the spectral
curve in this paper. Over ${\cal C}_{\vec{h}}$,
we have the Baxter vacuum state defined by the
following family of vectors in $
\stackrel{L}{\otimes} \CZ^N $: 
$$
|p\rangle \ : = |p_0\rangle \otimes \ldots \otimes
|p_{L-1}\rangle 
\in \ \ \stackrel{L}{\otimes} \CZ^N \ \ \ \ \ ( p
\in {\cal C}_{\vec{h}} ) \ ,
$$ 
where $| p_j \rangle$ is the vector in $\CZ^N$
governed by the relations: 
\be
\langle 0| p_j \rangle = 1 \ , \ \ \ \
\frac{\langle m|p_j \rangle}{\langle
m-1|p_j \rangle} = 
\frac{\xi_{j+1} a_j
\omega^m  - xb_j }{
- \xi_j (  \xi_{j+1} x c_j \omega^m - d_j ) }  \ .
\ele(Bv)
The constraint of the variables $(x, \vec{\xi})$
in 
${\cal C}_{\vec{h}}$ ensures that the following  
properties  hold for the Baxter vacuum state,
\begin{eqnarray*}
\widetilde{L}_{\vec{h}; 1 1}(x,
\vec{\xi})|p\rangle =  |\tau_- p\rangle 
\Delta_-(p) , &
\widetilde{L}_{\vec{h}; 2 2}(x,
\vec{\xi})|p\rangle =  |\tau_+
p\rangle\Delta_+(p) \ , &
\widetilde{L}_{\vec{h}; 2 1}(x,
\vec{\xi})|p\rangle = 0
\ , 
\end{eqnarray*}
where $\Delta_\pm, \tau_\pm$ are (rational)
functions and  automorphisms of 
${\cal
C}_{\vec{h}}$, defined by
\begin{eqnarray}
&\Delta_-(x, \xi_0, \ldots, \xi_{L-1})  &=
\prod_{j=0}^{L-1}( d_j-x
\xi_{j+1} c_j ) \ , \nonumber \\
&\Delta_+(x, \xi_0, \ldots, \xi_{L-1}) &= 
\prod_{j=0}^{L-1} \frac{\xi_j
(a_jd_j-x^2b_jc_j)}{\xi_{j+1}a_j -xb_j} \  , 
\label{DelTau} \\
&\tau_\pm :  (x, \xi_0, \ldots, \xi_{L-1}) 
&\mapsto (q^{\pm 1} x, 
q^{-1} \xi_0, \ldots, q^{-1} \xi_{L-1}) \ 
\nonumber .
\end{eqnarray}
This implies that, under the action of 
the transfer matrix, the Baxter vacuum state is
the sum of its
$\tau_\pm$-translations:
\bea(l)
T_{\vec{h}}(x) |p \rangle = |\tau_- p\rangle 
\Delta_-(p)  + |\tau_+ p\rangle\Delta_+(p) \ , \ \ 
{\rm for } \ \ p \in {\cal C}_{\vec{h}} \ .
\elea(T|p)
For a common eigenvector $\langle \varphi|
\in \stackrel{L}{\otimes}\CZ^{N*}$ of
$T_{\vec{h}}(x) \ (x \in \CZ) $, its
eigenvalue
$\Lambda(x)$ is a polynomial of $x$, i.e., 
$\Lambda(x) \in \CZ[x]$; while its values on the
Baxter's vacuum state, $Q(p):= \langle 
\varphi|p\rangle\  (p \in
{\cal C}_{\vec{h}})$, is a rational function on
${\cal C}_{\vec{h}}$.  Then the
following relation hold for $\Lambda (x)$ and
$Q(p)$:
\bea(l)
\Lambda(x) Q(p)  = \Delta_-(p) Q(\tau_-(p))  
+ \Delta_+(p) Q(\tau_+(p))  \ , \ \ {\rm for} \ p 
\in {\cal C}_{\vec{h}} \ ,
\elea(Bethe)
which will be called the  Baxter's $T$-$Q$
relation  for $T_{\vec{h}}(x)$
Note that the $\langle
\varphi|$ is also a common eigenvector of
$T^\ast_{\vec{h}}(x)$  with the eigenvalue
$\Lambda^\ast(x)= q^n \Lambda(x)$, where $q^n$ is
the $D^{\frac{-1}{2}}$-eigenvalue of $\langle 
\varphi|$. The Baxter's $T$-$Q$ relation for
$T^\ast_{\vec{h}}(x)$ becomes
\bea(l)
\Lambda^\ast(x) Q(p)  = \Delta^\ast_-(p)
Q(\tau_-(p))   + \Delta_+(p)^\ast Q(\tau_+(p))  \
,
\
\ {\rm for} \ p 
\in {\cal C}_{\vec{h}} \ ,
\elea(BetheT*)
with $\Lambda^\ast(x)= q^n \Lambda(x)$ , $ 
\Delta^\ast_-(p)= q^n \Delta_-(p)$, and $
\Delta_+(p)^\ast = q^n
\Delta_+(p) $ for $n \in \ZZ_N$. 
For $L$=4, one has
\be
T_{\vec{h}} (x) = T_0 
+ x^2 T_2 + x^4 T_4 \ , \ \ \ \ \ 
T^{\ast}_{\vec{h}} (x) = T^{\ast}_0 
+ x^2 T^{\ast}_2 + x^4 T^{\ast}_4 \ ,
\ele(TT*)
where $T_{2j}$ are operators of $
\stackrel{4}{\otimes} \CZ^N$ with the
expressions:
$$
\begin{array}{ll}
T_0 &= a_0a_1a_2a_3 \omega^2 D  + d_0d_1d_2
d_3 \ , \\
 T_2 &= a_0a_1b_2c_3 Y \otimes Y \otimes X
\otimes Z  + b_0c_1 a_2 a_3 X \otimes Z \otimes Y
\otimes Y  + a_0 b_1 c_2a_3 Y \otimes X \otimes Z
\otimes Y \\
 & + a_0b_1
d_2c_3 Y \otimes X \otimes1 \otimes Z + b_0d_1
c_2a_3 X \otimes 1 \otimes Z \otimes Y + b_0d_1d_2c_3 X \otimes 1 
\otimes 1 \otimes Z \\
 & + (a_jY \leftrightarrow d_j ,
 b_j X \leftrightarrow  c_j Z ) \ , \\
 T_4 &= b_0c_1b_2c_3  D C^{-1}
+ c_0b_1 c_2b_3 C \ , \ \ C:= Z \otimes X \otimes
Z \otimes X \ .
\end{array}
$$
One can also obtain the
expressions of
$T^{\ast}_{2j}$s from (\req(TT*r)). Note that 
the operators $C, D$ 
$T_2$ and $T_2^\ast$ commute each other. 

For
the study of discrete quantum
pendulum and SG
model in this
paper, we shall restrict ourselves on the 
case $L$=4 with the following
constraint\footnote{The convention we use here is
in tune with the one in
\cite{BKP}.} in 
$\vec{h}$,
\begin{eqnarray}
a_jd_j=q^{-1} , \ b_jc_j=-k^{-1} \ ,  & {\rm
for} \ \ \ j=0, 2 \ , \nonumber
\\
a_jd_j=q^{-1} , \ b_jc_j= -k \ , \ \ \ & {\rm for}
\
\ \ j =1, 3 \ . \label{condPSC}
\end{eqnarray}
where $k$ is a complex parameter.
Then the operators $T_{2j}, T_{2j}^\ast$
in (\req(TT*)) now take the following
forms:
\bea(ll)
\ \ T_0 = &\frac{1}{d_0d_1d_2d_3} D + 
d_0d_1d_2d_3 \ , \ \ \ \ \ \ \ \ \ \ \ \ \ \ \ \ \
\ T_4
 = \ \frac{c_1c_3}{k^2c_0c_2}  D C^{-1} +
\frac{k^2c_0c_2}{c_1c_3}C \ ,  \\
-T_2 =& \frac{  kc_0 d_2d_3 }{ c_1}  U_1 +
\frac{ d_0 c_1 d_3}{kc_2 } U_2 +
\frac{k d_0d_1 c_2 }{c_3} U_3 
  + 
\frac{ d_1d_2c_3 }{k c_0 } U_4 
+  \frac{ c_1}{  kc_0 d_2d_3 }  D U_1^{-1} 
+ \frac{kc_2 }{ d_0 c_1 d_3} D U_2^{-1} 
 \\
& + \frac{c_3}{k d_0d_1 c_2 } D U_3^{-1} 
 + \frac{k c_0 }{ d_1d_2c_3 } D  U_4^{-1} +  
\frac{ kd_0c_1}{qd_2c_3} V_1 + \frac{k
qd_2c_3}{d_0c_1}  D V_1^{-1} +   
\frac{ d_3
c_0}{kqd_1c_2}V_4 +\frac{q   
d_1c_2}{kd_3c_0} DV_4^{-1}  \ ; \\ [2mm]
\ \ T^{\ast}_0 = & \frac{1}{d_0d_1d_2d_3}
D^{\frac{1}{2}} +  d_0d_1d_2d_3 D^{\frac{-1}{2}}
\ , \ \ \ \ \ \ \ \ \ \ \ 
T^{\ast}_4  =  \  
 \frac{c_1c_3}{k^2c_0c_2}  
D^{\frac{1}{2}} C^{-1} + 
\frac{k^2c_0c_2}{c_1c_3}D^{\frac{-1}{2}}C \ ,
\\ -T^{\ast}_2 = &
\frac{  kc_0 d_2d_3 }{ c_1}   D^{\frac{-1}{2}}U_1
+
\frac{ d_0 c_1 d_3}{kc_2 } D^{\frac{-1}{2}}U_2 +
\frac{k d_0d_1 c_2 }{c_3} D^{\frac{-1}{2}}U_3 
  + 
\frac{ d_1d_2c_3 }{k c_0 }D^{\frac{-1}{2}} U_4 
+  \frac{ c_1}{  kc_0 d_2d_3 }  D^{\frac{1}{2}} U_1^{-1} 
\\& + \frac{kc_2 }{ d_0 c_1 d_3} 
D^{\frac{1}{2}} U_2^{-1}  + \frac{c_3}{k
d_0d_1 c_2 } D^{\frac{1}{2}} U_3^{-1} 
 + \frac{k c_0 }{ d_1d_2c_3 } D^{\frac{1}{2}}
U_4^{-1} 
+  \frac{k
d_0c_1}{qd_2c_3} D^{\frac{-1}{2}}V_1 + \frac{
k qd_2c_3}{d_0c_1}  
D^{\frac{1}{2}} V_1^{-1} \\& + 
\frac{ d_3
c_0}{kqd_1c_2}D^{\frac{-1}{2}} V_4 +\frac{q   
d_1c_2}{kd_3c_0} D^{\frac{1}{2}} V_4^{-1}   \ ,
\elea(TTsp)
where $U_j, V_j$ are operators defined by 
$$
\begin{array}{llll}
U_1 = 
Z \otimes X \otimes 1 \otimes 1 , & 
 U_2 = 1 \otimes Z \otimes X \otimes 1 ,&  
U_3= 1 \otimes 1 \otimes Z \otimes X , & U_4 = 
X \otimes 1 \otimes 1 \otimes Z , \\
V_1 = 1 \otimes Z \otimes Y \otimes X , & V_2 = 
X \otimes 1 \otimes Z \otimes Y , & V_3 = 
Y \otimes X \otimes 1 \otimes Z , & V_4 = Z
\otimes Y \otimes X  \otimes 1 \ .
\end{array} 
$$
It is easy to see that the following relations
hold among the above operators,
\bea(ll)
U_{j+1}U_j = \omega U_jU_{j+1} \ , & V_{j+1}V_j =
\omega^2 V_jV_{j+1} , \ \ \  (U_5:=U_1, V_5:=V_1)
,
\\  U_i U_j =
U_jU_i
\ , & V_i V_j =
V_jV_i  \ \ {\rm if } \ i \equiv j
\pmod{2} ;  \\
V_1 =  U_3U_2 \ , \ V_2 = 
U_4U_3 ,  & V_3= U_1U_4 \ , \ V_4= U_2U_1
\ ;
\\  U_1U_3 = C \ , \ U_2U_4 = C^{-1}D \ , & 
V_1V_3=V_2V_4=
\omega D 
 \ .
\elea(UV)
Under the constraint (\ref{condPSC}), the
functions 
$\Delta_\pm$ in (\ref{DelTau})  become 
$$
\begin{array}{ll}
\Delta_-(x, \xi_0, \ldots, \xi_3 )  &=
\prod_{j=0}^3 ( d_j-x
\xi_{j+1} c_j ) \ ,  \\ [2mm]
\Delta_+(x, \xi_0, \ldots, \xi_3) &= 
 \frac{d_0d_1d_2d_3 (1 +x^2qk^{-1})^2
(1 +x^2qk)^2}{(1 +x 
\xi_1^{-1}d_0c^{-1}_0q k^{-1})(1 +x 
\xi_3^{-1}d_2c^{-1}_2q k^{-1})(1 +x 
\xi_2^{-1}d_1c^{-1}_1q k)(1 +x 
\xi_0^{-1}d_3c^{-1}_3q k) } \ .
\end{array}
$$
By (\req(Bv)), the Baxter vacuum
state $|p \rangle =\otimes_{j=0}^3 |p_j\rangle $
have the following expression, 
$$
\langle m|p_j \rangle = \left\{ \begin{array}{ll}
\frac{ \xi_{j+1}^mq^{m^2}
(- k^{-1} x 
\xi_{j+1}^{-1}c_j^{-1}d_jq^{-1} ; \ \omega^{-1} )_m }{ \xi_j^m d^{2m}(
x \xi_{j+1}c_jd_j^{-1}q^2 ; \ \omega)_m} & {\rm
for \ even } \ j , \\
\frac{ \xi_{j+1}^mq^{m^2}
(- k x 
\xi_{j+1}^{-1}c_j^{-1}d_jq^{-1} ; \ \omega^{-1} )_m }{ \xi_j^m 
d_j^{2m}(x \xi_{j+1}c_jd_j^{-1}q^2 ; \ \omega)_m}
& {\rm
for \ odd } \ j \ .
\end{array} \right.
$$
In this paper we shall mainly study the
diagonalization problem of $T_2^\ast$, or
equivalently 
$T_2$ , under the condition 
(\ref{condPSC}), plus the following further
constraints on parameters $d_j, c_j$ and the
operator $C$:
\be
d_0d_1d_2d_3= 1 \ , \ c_1^Nc_3^N = k^{2N}
c_0^Nc_2^N \ , \ \ C = \frac{c_1c_3}{k^2c_0c_2} 
\ .
\ele(DQPSG)
The above properties on the parameters
arise from the connection of
$T_2^\ast$ with  the following
physical models.

(I) Discrete quantum pendulum. This is the
situation under the constraint (\req(DQPSG)),
and with the further identifications:
$$
D=1 \ ,  \ \ \ 
 \frac{ d_0 c_1 d_3}{kc_2 } U_2 = 
\frac{k c_0 }{ d_1d_2c_3 }   U_4^{-1} ( = :
Q_{n-1}) \ , \ \ \frac{ c_1}{  kc_0 d_2d_3 }  
U_1^{-1}  = 
\frac{k d_0d_1 c_2 }{c_3} U_3 (=:Q_n )  .
$$
By (\req(UV)), one has  
$$
Q_{n-1}Q_n= \frac{ d_0 c_1}{   d_2c_3 }
\omega^{-1} V_1 = 
 \frac{ d_0 c_1}{   d_2c_3 }  V_3^{-1} \ , \ \ \ 
Q_{n-1}Q_n^{-1} = \frac{ c_0d_3 }{ c_2d_1 }   V_4 = 
\frac{ c_0d_3 }{ c_2d_1 } \omega V_2^{-1} \ .
$$
Then $T^\ast_{2j}$ in (\req(TTsp)) become
\bea(ll)
\ \ T_0^\ast &= T_4^\ast = 2 \\ [1mm]
-T_2^\ast &= 
2(Q_n +Q_n^{-1} + Q_{n-1}
  + 
Q_{n-1}^{-1}) \\ &
+  k ( qQ_{n-1}Q_n + q^{-1} Q_n^{-1}Q_{n-1}^{-1}) + 
 k^{-1}(   qQ_nQ_{n-1}^{-1}+ q^{-1} Q_{n-1}Q_n^{-1}
) \ .
\elea(DQP)
The above $-T_2^\ast$ is the
Hamiltonian of discrete quantum pendulum
in \cite{BKP}, subject to the following
evolution equation: 
$$
Q_{n+1}Q_{n-1} = (\frac{k+qQ_n}{1+qkQ_n})^2 \ , \
\  Q_nQ_{n-1}= q^2 Q_{n-1}Q_n \ .
$$

(II) Discrete sine-Gordon (SG) Hamiltonian. This
is the situation under the constraint (\req(DQPSG))
with one further identification:
$$
\frac{c_1d_0d_3}{kc_2} D^{\frac{-1}{2}} U_2 =
\frac{kc_0}{c_3d_1d_2} D^{\frac{1}{2}} U_4^{-1} \
\ .
$$
In this case, we have 
\bea(ll)
T^{\ast}_0 = & 
D^{\frac{1}{2}} +  D^{\frac{-1}{2}}
\ , \ \ \ \ \ \ \ \ \ \ \ 
T^{\ast}_4  =  \  
D^{\frac{-1}{2}} +  D^{\frac{1}{2}} \ ,
\\ [1mm]
-T^{\ast}_2 = &
\frac{  kc_0 d_2d_3 }{ c_1}   D^{\frac{-1}{2}}U_1
+
\frac{ c_1d_0  d_3}{kc_2 } D^{\frac{-1}{2}}U_2 +
\frac{k  c_2 d_0d_1}{c_3} D^{\frac{-1}{2}}U_3 
  + 
\frac{ c_3d_1d_2 }{k c_0 }D^{\frac{-1}{2}} U_4 
+  \frac{ c_1}{  kc_0 d_2d_3 }  D^{\frac{1}{2}} U_1^{-1} 
\\& + \frac{kc_2 }{  c_1 d_0 d_3} 
D^{\frac{1}{2}} U_2^{-1}  + \frac{c_3}{k
c_2 d_0d_1 } D^{\frac{1}{2}} U_3^{-1} 
 + \frac{k c_0 }{ c_3 d_1d_2 } D^{\frac{1}{2}}
U_4^{-1} 
+  \frac{k
c_1d_0}{qc_3d_2} D^{\frac{-1}{2}}V_1 + \frac{
k qc_3d_2}{c_1d_0}  
D^{\frac{1}{2}} V_1^{-1} \\& + 
\frac{c_0 d_3
}{kqc_2d_1}D^{\frac{-1}{2}} V_4 +\frac{q   
c_2d_1}{kc_0d_3} D^{\frac{1}{2}} V_4^{-1}   \ .
\elea(SG)
The above $-T_2^*$ can be identified with
the  discrete quantum sine-Gordon integral in
\cite{BKP}, for which a
detailed description will be given in the appendix
of this paper.

\section{The Baxter's  T-Q Polynomial Equation for
Rational Degenerated Spectral Curve } In this
section,
 we derive the  Baxter's
$T$-$Q$ polynomial relation for a 
size
$L$ when the spectral curve ${\cal C}_{\vec{h}}$
is degenerated into rational curves, i.e., 
${\cal C}_{\vec{h}}$  is a disjoint union of
finite copies of the base 
$x$-curve.  We shall reduce the general
degenerated situation  to one special case,  
which we
have already discussed in our previous article
\cite{LR}. 

By the rational degenerated spectral
curves, we mean the   
coordinates $\xi_j^N$ of 
${\cal C}_{\vec{h}}$ to be
constants, i.e., independent of the variable $x$
for all $j$.  Then the parameters
$h_j$ and the variables $\xi_j$ are subject
to the relations:
$$
\frac{b_j^Nd_j^N}{a_j^Nc_j^N} = \frac{a_{j+1}^Nb_{j+1}^N}{c_
{j+1}^Nd_{j+1}^N} \ , \ \ \ \
\xi_j^{2N} = \frac{a_j^Nb_j^N}{c_j^Nd_j^N} \ \ \
\ \ \ \ {\rm for} \ \ 0 \leq j \leq L-1 \ .
$$
In this situation, we define
\be
r_j  =
\sqrt{\frac{b_{j-1}d_{j-1}}{a_{j-1}c_{j-1}}} \ ,
\  \  \ \ \ j \in \ZZ_L \ .
\ele(rj)
Then ${\cal C}_{\vec{h}}$ contains the following 
$\tau_\pm$-invariant curve ${\cal C}$, over which
we shall formulate the
Baxter's
$T$-$Q$ equation,
$$
{\cal C} := \{ (x, \xi_0, \ldots, \xi_{L-1}) \ | \ 
 r_0^{-1}\xi_0 = \ldots = 
r_{L-1}^{-1}\xi_{L-1} =  q^l \ , \ \ l \in \ZZ_N
\} \ .
$$ 
We shall make the identification of ${\cal C}$ with $\PZ^1
\times \ZZ_N$ via the following correspondence:
$$
{\cal C} = \PZ^1 \times \ZZ_N \ , \ \ \ \ 
(x, \  r_0 q^l, \ldots, 
r_{L-1} q^l) \longleftrightarrow 
(x, l) \ .
$$
Then the automorphisms $\tau_\pm$ on ${\cal C}$  
are expressed by 
$$
\tau_\pm :  \ (x, l) \mapsto (q^{\pm 1} x, l-1) \ ,
$$
and the action  $T(x) \ (:=
T_{\vec{h}}(x))$ on $|x, l\rangle$ in (\req(T|p))
now takes the form: 
\bea(l)
T (x) |x, l\rangle = |q^{-1}x, l-1\rangle 
\Delta_-(x, l)  + |qx, l-1\rangle\Delta_+(x, l) \  , 
\elea(T|xm)
where $\Delta_\pm$ are the following rational
functions of $x$: 
\begin{eqnarray*}
&\Delta_-(x, l)  &= ( d_0 \cdots d_{L-1} ) 
\prod_{j=0}^{L-1}( 1-x
  q^l d_j^{-1}c_j r_{j+1} ) \ ,\\
&\Delta_+(x, l) &= ( d_0 \cdots d_{L-1} ) 
\prod_{j=0}^{L-1} \frac{
1-x^2a_j^{-1}d_j^{-1}b_jc_j}{1 -x q^{-l} a_j^{-1}b_j
r_{j+1}^{-1}} 
\  . 
\end{eqnarray*}
With the substitutions,
$$
(d_0\cdots d_{L-1})^{-1} T(x) \mapsto T(x) \ , \ \ 
(d_0\cdots d_{L-1})^{-1}\Delta_\pm(x, l) \mapsto \Delta_\pm (x, l ) 
\ ,
$$
the relation (\req(T|xm)) still holds for the
modified
$\Delta_\pm$, now with the expressions: 
$$
\Delta_-(x, l)  = 
\prod_{j=0}^{L-1}( 1-x
   c^*_j q^l) \ , \ \ \ \
\Delta_+(x, l) = 
\prod_{j=0}^{L-1} \frac{
1-x^2c^{* 2}_j}{1 -x c^*_j q^{-l} } \  ,
$$
where 
\be
c^*_j := d_j^{-1}c_j r_{j+1} \ \ 
(= a_j^{-1}b_j r_{j+1}^{-1}) \ . 
\ele(c*j)
Furthermore,  one can convert the expression
(\req(Bv)) of the Baxter vacuum state over ${\cal
C}$ to the following component-expression of the
Baxter's vector $|x, l \rangle$:
\begin{eqnarray*}
\langle {\bf k}|x, l \rangle  = q^{|{\bf k}|^2}
\prod_{j=0}^{L-1} 
\frac{ (x c^*_jq^{-l-2}; \omega^{-1})_{k_j} }{ (x 
c^*_j q^{l+2}; \omega )_{k_j}}
 \ . 
\end{eqnarray*}
Here the bold letter ${\bf k}$ denotes a
multi-index vector
${\bf k}= (k_0, \ldots, k_{L-1})$ for 
$k_j \in \ZZ_N$, and the square-length of
${\bf k}$ is defined by 
$|{\bf k}|^2:= \sum_{j=0}^{L-1} k_j^2$.  
Each ratio-term  in the above right
hand side is given by a non-negative 
representative for each element in $\ZZ_N$ 
appeared in the formula. With the above
description of $T(x)$ on the Baxter
vacuum state
$|x, l\rangle$, the discussions of  Sect. 4 and 
Sect. 5  Proposition 2, 3  in
\cite{LR}  can
be applied to our present situation. This enables
us to state the following  result on the
Baxter's
$T$-$Q$ equation and its  connection with the
transfer matrix 
$T(x)$:
\begin{theorem}  \label{thm:LR}
 Let $f^e, f^o$ be functions on 
${\cal C}$ defined by
\begin{eqnarray*}
f^e(x, 2n) =  \prod_{j=0}^{L-1}
\frac{(xc^*_j ; \omega^{-1} )_{n+1} }{
(xc^*_j; \omega )_{n+1}} , 
\ \ \
f^o(x, 2n+1) =  \prod_{j=0}^{L-1}
\frac{(xc^*_jq^{-1}; \omega^{-1} )_{n+1}}{
(xc^*_jq ; \omega)_{n+1}} \ .
\end{eqnarray*}
For $x \in \PZ^1$ and $ l \in \ZZ_N$,
we define the following vectors in
$\stackrel{L}{\otimes}\CZ^N$, 
\begin{eqnarray*}
|x\rangle_l^e = \sum_{n=0}^{N-1}  |x, 2n\rangle  
f^e (x, 2n)
\omega^{ln} 
 \ , &
|x\rangle_l^o = \sum_{n=0}^{N-1} 
 |x, 2n+1\rangle f^o (x, 2n+1)  \omega^{ln}  \ , 
\\
|x \rangle_l^+ =  |x\rangle_l^e q^{-l} u(qx) +  
|x \rangle_l^o u(x) & {\rm where} \ \ 
u(x): = \prod_{j=0}^{L-1} (1-x^Nc^{*
N}_j)(xc^*_jq ; q^2)_M \ .
\end{eqnarray*} 
Then 

(i) $|x \rangle_l^e u(qx) = 
|x\rangle_l^o q^lu(x)$, or equivalently, 
$
|x\rangle_l^+ = 2 q^{-l} |x\rangle_l^e u(qx)  = 2
|x\rangle_l^ou(x) $. \\[.1mm]

(ii) The $T(x)$-transform on $| x\rangle^+_l$ is
given by
$$
q^{-l} T(x)|x\rangle^+_l
=|q^{-1}x\rangle^+_l  \Delta_-(x, -1) +
|qx\rangle^+_l  \Delta_+(x, 0) \ , \ \ \ l \in \ZZ_N \ .
$$

(iii) For a common eigenvector $\langle \varphi|$
of
$T(x)$ with the  eigenvalue
$\Lambda (x)$, the function
$Q_l^+(x)  (:=\langle \varphi | x\rangle^+_l)$
and $\Lambda (x)$ are  polynomials with the
properties:
$$
\begin{array}{llll}
{\rm deg} Q^+_l (x)
\leq  (3 M +1)L , &
 {\rm deg}
\Lambda (x) \leq 2[\frac{L}{2}], &
\Lambda (x)= \Lambda(-x), & 
\Lambda (0)= 
q^{2l}+1  ,
\end{array}
$$
and they satisfy the following Baxter's $T$-$Q$
equation:
\begin{eqnarray}
 q^{-l} \Lambda (x) Q_l^+ (x) = 
\prod_{j=0}^{L-1}(1-xc^*_jq^{-1})
Q_l^+ ( xq^{-1}) + \prod_{j=0}^{L-1} 
(1+x c^*_j) Q_l^+ (xq) \ .
\label{QmEq}
\end{eqnarray}
Furthermore, for $0 \leq m \leq M$, 
$Q_m^+(x)$ and $ Q_{N-m}^+(x)$ are elements in 
$ x^m \prod_{j=0}^{L-1} (1-x^Nc^{* N}_j)
\CZ[x]$. 
\end{theorem}
$\Box$ \par \vspace{0.2in} \noindent
For the rest of this paper, the letter $m$ will
always denote an integer between 
$0$ and $M$,
$$
0 \leq m \leq M \ . 
$$
By $(iii)$ of the Theorem \ref{thm:LR},
the  equation (\ref{QmEq}) for the sectors $m,
N-m$ can be combined into a single one by
introducing the  polynomials
$\Lambda_m(x), Q ( x)$ via the relation,
$$
(\Lambda_m(x) , \ x^m
\prod_{j=0}^{L-1} (1-x^Nc^{* N}_j) Q ( x) ) = 
(q^{-m}\Lambda(x), \ Q_m^+(x) ) , \ \ (q^m
\Lambda (x) , \ Q_{N-m}^+(x) )  \ .
$$
Then the equations (\ref{QmEq}) for $l=m, N-m$ 
are equivalent to the  following 
polynomial equation of $Q (x), \Lambda_m(x)$: 
\be
 \Lambda_m (x) 
Q (x) = q^{-m} 
\prod_{j=0}^{L-1}(1-xc^*_jq^{-1}) 
Q (xq^{-1})+ q^m \prod_{j=0}^{L-1} (1+x c^*_j) 
 Q (xq) \ ,
\ele(rBeq)
with the following constraints of $Q(x)$ and $
\Lambda_m(x)$,
$$
{\rm deg} \ Q (x)
\leq ML-m ,  \ \ 
 {\rm deg} \
\Lambda_m (x) \leq 2[\frac{L}{2}], \ \ 
\Lambda_m (x)= \Lambda_m (-x),  \ \ \Lambda_m
(0)=  q^m + q^{-m}. 
$$
By (\req(TT*r)), the above $\Lambda_m(x)$ is
indeed  the eigenvalue of $T^{\ast}(x) $; while
(\req(rBeq)) corresponds the Baxter's $T$-$Q$
equation (\req(BetheT*)) for 
$T^{\ast}(x)$ on the sectors $m, N-m$.

For $L$ even, by the construction of
$T^{\ast}_{\vec{h}}(x)$, one can see that the
eigenvalues of $T_L^\ast$ are non-zero,
hence ${\rm deg} \ \Lambda_m(x)= L$. In the
situations we will consider later on, the polynomial solutions,
$Q(x), \Lambda_m(x)$ of the
equation (\req(rBeq)) possesses certain
reciprocal symmetry property. Here we call a
polynomial
$P(x)$ to be reciprocal if
$P^\dagger(x)=P(x)$, where $P^\dagger(x)$ is the
polynomial defined by
$$
P^\dagger (x) := x^{{\rm deg} P}P(x^{-1}) \ .
$$
\begin{proposition} \label{prop:syTQ}
For $L$ even, assume that the
polynomial $\Lambda_m(x)$  in $(\ref{QmEq})$ is 
reciprocal, and the parameters
$c_j^*$s and the degree $d$ of $Q(x)$ satisfy
the following properties:

(i) $\{ c_0^*, \ldots, c_{L-1}^* \} = \{ -c_0^{*
-1}q , \ldots, -c_{L-1}^{* -1}q \}$ .

(ii) $\prod_{j=0}^{L-1} c^*_j =
q^{\frac{L}{2}}, \  \ \ \ q^{d+2m + \frac{L}{2}}=1
$.
\par \noindent 
Then $Q^\dagger(x)$ is also a
solution of
$(\ref{QmEq})$ for $\Lambda_m(x)$.
\end{proposition}
{\it Proof.} 
By substituting $x$ by $x^{-1}$ in (\ref{QmEq}),
and then multiplying $x^{d+L}$ to the equation,
one obtains the relation,
\begin{eqnarray*}
 \Lambda_m (x) 
Q^\dagger (x) = q^{m+d} \prod_{j=0}^{L-1} c^*_j
\prod_{j=0}^{L-1}(1 +xc_j^{* -1})  
 Q^\dagger (xq^{-1}) +q^{-m-d-L} 
\prod_{j=0}^{L-1} c^*_j
\prod_{j=0}^{L-1}(1-xc_j^{* -1}q) 
Q^\dagger(xq) \ ,
\end{eqnarray*}
By  $(ii)$, we have
$$
 q^{m+d} \prod_{j=0}^{L-1} c^*_j = q^{-m} \ , \
\ \ \  q^{-m-d-L} 
\prod_{j=0}^{L-1} c^*_j = q^m \ .
$$
Then, by $(i)$, the above equation of
$Q^\dagger(x)$ is the same as (\req(rBeq)). 
$\Box$ \par \vspace{0.2in} \noindent

The following algebraic fact was shown  in
\cite{LR} Lemma 6, which  we just state here
for later use. 
\begin{lemma}\label{lem:det} 
Let $n$ be an odd positive integer, $A$ be a $n
\times n$-matrix with complex entries $a_{i,j}$
satisfying the relations
$$
a_{i,j}= (-1)^{i+j+1} a_{n-j+1, n-i+1} \ , \ \ 
{\rm for } \ 1 \leq i, j \leq n \ .
$$
Then  $A$ is a degenerated matrix. 
\end{lemma}
$\Box$

\section{The Baxter's T-Q Polynomial Relation for
L=4}  
For $L$=4  rational
degenerated case, the parameter $\vec{h}$
we  discuss later in this paper will subject
to the constraint (\ref{condPSC}), and be confined
only to the following situation:
\be
qa_j= d_j= 1 \ , \ \ \ 
-b_j=c_j= \left\{\begin{array}{ll}
k^{\frac{-1}{2}}& {\rm for \ even} \ j , \\
k^{\frac{1}{2}} &{\rm for \ odd} \ j .
\end{array}
\right.
\ele(deg4)
Then, by (\req(rj)) and (\req(c*j)) we
have 
$r_j= (-q)^{\frac{1}{2}}$ for all $j$ ,
and 
\be
c^*_j =
\left\{\begin{array}{ll}
(-q)^{\frac{1}{2}} k^{\frac{-1}{2}}& {\rm
for \ even}
\ j , \\
(-q)^{\frac{1}{2}} k^{\frac{1}{2}}
&{\rm for \ odd} \ j .
\end{array}
\right.
\ele(c*)
The operators (\req(TTsp)) in 
$T_{\vec{h}}^{\ast}(x)$  become
$$
\begin{array}{rl}
T_0^{\ast} &= D^{\frac{1}{2}} +  
D^{\frac{-1}{2}}, \ \ \ \ \ \ \ \ \ \ \ 
T_4^{\ast} \ =    D^{\frac{1}{2}} C^{-1}
+D^{\frac{-1}{2}} C \ , 
\\  -T_2^{\ast} &= D^{\frac{-1}{2}}U_1
+ D^{\frac{-1}{2}}U_2 + D^{\frac{-1}{2}}U_3 
  + D^{\frac{-1}{2}} U_4 + k q^{-1}
D^{\frac{-1}{2}}V_1 + 
 k^{-1}q^{-1}D^{\frac{-1}{2}} V_4
\\ &
+   D^{\frac{1}{2}}
U_1^{-1}  + 
D^{\frac{1}{2}} U_2^{-1} + D^{\frac{1}{2}}
U_3^{-1} 
 +  D^{\frac{1}{2}} 
U_4^{-1} 
+  k
q    D^{\frac{1}{2}}
V_1^{-1} + 
 k^{-1} q D^{\frac{1}{2}}
V_4^{-1}  \ .
\end{array}
$$
By $C^N=1$, the polynomial 
$\Lambda_m(x)$ in (\req(rBeq)) now takes the
form:
\be
\Lambda_{m, l}(x) = (q^{m+l}+q^{-m-l}) x^4 +
\lambda x^2 + q^m + q^{-m} \ , \ \ \ \ 
0 \leq m \leq M, \ \ 0 \leq l \leq 2M \ ,
\ele(lambda)
where $\lambda$ is an eigenvalue of $-T^\ast_2$
for the sectors $(m, l),  (N-m, l)$, which can
be regarded as the label of the eigenvalues
of
$D^{\frac{\pm1}{2}}$ and $ C^{\mp 1}$. The
Baxter's
$T$-$Q$ polynomial equation (\req(rBeq)) 
now takes the form,
\begin{eqnarray}
 \Lambda_{m,l} (x) 
Q (x) = q^{-m} \triangle( x (- q)^{\frac{-1}{2}})
Q (xq^{-1})+ q^m \triangle(x (-
q)^{\frac{1}{2}}) 
 Q (xq) \ , \label{BEml}
\end{eqnarray}
with $$
\triangle(x) = (1 + 2 c x + x^2)^2 
\ , \ \ \ \ c := \frac{1}{2}( k^{\frac{1}{2}}+
k^{\frac{-1}{2}})  \ , 
$$ 
and  ${\rm deg} \ Q(x) \leq 4M-m$. The polynomials
$\Lambda_{m,l}(x) , Q(x)$ will be called the
eigenvalue and the eigen-polynomial of
(\ref{BEml}) respectively, whenever
$Q(x)$ is a non-trivial function. Indeed, the 
value $\lambda$ in the
expression of $\Lambda_{m,l}(x)$ is an eigenvalue
of $-T_2^{\ast}$. In the following, we are going
to study the equation (\ref{BEml}) for a generic
$\lambda$ in (\req(lambda)).

For the rest of this paper, the parameter
$c$ will always  be a generic complex number
unless otherwise stated. The polynomial
$Q(x)$  will  be denoted 
 by 
$$
 Q (x) = \sum_{j=0}^d \alpha_j x^j \ , \ \ \ d:=
{\rm deg} \ Q (x) \ \ ,
$$
and we define $\alpha_j:=0$ for 
$j$ not between $0$ and $d$. An equivalent
formulation of the equation (\ref{BEml})
is  the following system of difference equations
in $\lambda$ and $\alpha_j$s :
\be
\nu_j \alpha_j +
 v_j 
\alpha_{j-1} +
(\delta_j - \lambda) \alpha_{j-2} +u_j
\alpha_{j-3} +  \mu_j
\alpha_{j-4} = 0 \ , \ \  \  ( j \geq 1 ) \ , 
\ele(eqDL4)
where the coefficients in the above equations are
defined by
\bea(ll)
\nu_j = q^{m+j}+q^{-m-j}-q^m-q^{-m} , &
v_j = 
4c{\rm
i}(q^{m+j-\frac{1}{2}}-q^{-m-j+\frac{1}{2}}) ,
\\
\delta_j = - (4c^2+2)(q^{m+j-1} +q^{-m-j+1} ), &
\\ u_j  = - 4c {\rm i} (q^{m+j- \frac{3}{2}
}-q^{-m-j+
\frac{3}{2} }) , &
\mu_j =  q^{m+j-2}+q^{-m-j+2}- q^{m+l}-q^{-m-l}
\ .
\elea(coef5)
Indeed, for the system (\req(eqDL4)), it suffices
to consider those relations for the index
$j$  between $1$ and $ d+3$. Note
that the relations in (\req(eqDL4)) for $2 \leq j
\leq d+2$ give rise to the following eigenvalue
problem:
\be
\left\{ \left( \begin{array}{ccccccc}
\delta_{d+2} & u_{d+2} &\mu_{d+2}&0&\cdots  & 0 & 0 \\
v_{d+1}& \delta_{d+1} &u_{d+1}&\mu_{d+1}&\ddots  
& 0 &0 \\
\nu_d&v_d& \delta_d &u_d&\mu_d&\ddots   &0 \\
0& \ddots& \ddots&\ddots &\ddots  & \ddots &\vdots \\
\vdots& \ddots & \ddots & \ddots & \ddots & 
\ddots &0\\
\vdots& \ddots &\nu_4 &v_4 & \delta_4 
&u_4&\mu_4\\
\vdots& \ddots  & \ddots &\nu_3 &v_3 & \delta_3 
&u_3\\
0&  \cdots & & 0 &\nu_2& v_2 & \delta_2
\end{array} 
\right) - \lambda \right\}
 \left( \begin{array}{c}
\alpha_d\\
\alpha_{d-1} \\
\vdots \\
\vdots \\
\vdots \\
\vdots\\
\alpha_0
\end{array} 
\right) = \vec{0} \ \ .
\ele(Mform)
Hence for a solution of the system
(\req(eqDL4)), the $\lambda$ can be regarded as
an algebraic function of $c$, and it has a limit
as $c$ tends to some special value $c_0$.   
\begin{lemma} \label{lem:d0} For equation
$(\ref{BEml})$ with a given 
$c$ (no generic property required), the degree $d$
of
$Q(x)$  satisfies the following conditions,
$$
1 \leq d \leq 4M-m  \ , \ \ \ \ \ q^{d+2}= q^l \
{\rm or}
\  q^{-2m-l} \ , 
$$
and the zero-multiplicity of $Q(x)$ at the
origin is equal to one of
$0, N, N-2m, 2N-2m$.
\end{lemma}
{\it Proof.} The upper bound
$4M-m$ of
$d$ is given by the assumption of
(\ref{BEml}). If $d=0$, then
a non-zero constant is a solution $Q(x)$ of
(\ref{BEml}), and we have 
$$
 \Lambda_{m,l}(x) = q^{-m} \triangle( x (-
q)^{\frac{-1}{2}}) + q^m \triangle(x (-
q)^{\frac{1}{2}})
\ . 
$$
By the even-function property of $
\Lambda_{m,l}(x)$, the above relation implies 
$q^{2m+1}= q^{2m+3}=1$, hence $q^2=1$, a
contradiction to the odd assumption on the
integer
$N$. Therefore
$d \geq 1$. Comparing the coefficients of the
highest degree of $x$ in (\ref{BEml}), one has
$$
q^{m+l} + q^{-m-l} = q^{-m-2-d} + q^{m+2+d} \ .
$$
This implies  $q^{m+2+d}= q^{m+l}$ or
$q^{-m-l}$,  i.e., $q^{d+2}= q^l$ or $q^{-2m-l}
$.  Denote $r$ the zero-multiplicity of $Q(x)$ at
$x=0$. By comparing the coefficients of degree
$r$  in (\ref{BEml}),  we have
$$
q^m + q^{-m} = q^{-m-r}+ q^{m+r} \ ,
$$  
hence $r \equiv 0, -2m \pmod{N}$. Then  the
conclusion of $r$  follows from $d \leq 4M-m$.
$\Box$ \par \vspace{0.2in} \noindent
{\bf Remark.} For the results 
obtained later in this paper on certain special
cases, and also on the similar problem of size
$L=3$ in \cite{LR},  the solution
$Q(x)$ in (\req(rBeq)) always possesses the
property 
$Q(0) \neq 0$. In this situation, one can
 write
$$
Q(x) = \prod_{j=1}^d (x-\frac{1}{z_j}) \ , \ \ \ \
z_j \neq 0 \ . 
$$
Substituting $x = \frac{1}{z_j}$ in
(\ref{BEml}), one obtains the following
relations of $z_j$s :
\begin{eqnarray}
   q^{2m+2+d} (\frac{z_j^2 + 2 {\rm i}c
q^{\frac{1}{2}} z_j -  q}{ q z_j^2 - 2  {\rm i} c
q^{\frac{1}{2}}z_j
 - 1})^2 = \prod_{n \neq j, n=1}^d  \frac{z_n-q
z_j}{qz_n - z_j} \ , \ \ \ \ \ 
j = 1, \ldots , d \ ,
\label{rtBZ}
\end{eqnarray}
which is the  Bethe
ansatz  equation appeared in literature, e.g.
\cite{FK}. 
$\Box$ \par \vspace{0.2in} \noindent
A special case in the above setting happens
when  $\Lambda_{m, l}(x)$ in 
(\ref{BEml}) is a reciprocal polynomial. For the 
convenience, the Baxter's $T$-$Q$
relation  (\ref{BEml}) will be called
a symmetric $T$-$Q$ polynomial relation if the
following condition holds:
$$
 \Lambda_{m,l}^\dagger(x) =
\Lambda_{m,l}(x), 
\  \ 
{\rm  equivalently \ } \ \ q^l=1, \ q^{-2m} \ , \
\ i.e., \ \ l \equiv 0 , N-2m \ \pmod{N} \ .
$$
In this situation, (\req(lambda)) becomes
\be
\Lambda_{m,l}(x)= q^m+q^{-m} + \lambda x^2 +
(q^m+q^{-m}) x^4
\ ,
\ele(lmbasy)
and the coefficients (\req(coef5)) in the system
(\req(eqDL4)) have the following
form,
\bea(ll)
\nu_j = q^{m+j}+q^{-m-j}-q^m-q^{-m} , & 
v_j = 
4c{\rm
i}(q^{m+j-\frac{1}{2}}-q^{-m-j+\frac{1}{2}}) ,
\\
\delta_j = - (4c^2+2)(q^{m+j-1} +q^{-m-j+1} ) ,
&
\\ 
u_j  = - 4c {\rm i} (q^{m+j- \frac{3}{2}
}-q^{-m-j+
\frac{3}{2} }), &
\mu_j =  q^{m+j-2}+q^{-m-j+2}- q^{m}-q^{-m} \ .
\elea(coefsy)
Note that by the equalities, $ 
u_{j+1}=-v_j$ and $ \mu_{j+2}= \nu_j$,
the transport of the square matrix in 
(\req(Mform)) is unchanged after
substituting $c$ by  $-c$. Hence the
eigenvalue
$\lambda$ for the symmetric
$T$-$Q$  polynomial relation necessarily
becomes an algebraic function of $c^2$,
equivalently, the following property  holds for
$\lambda$:
\be
\lambda =
\lambda(c) = \lambda(-c) \ .
\ele(SyDef)
Furthermore,
the  relation (\ref{BEml}) is unchanged when
substituting  $(c, x)$ by $
(-c, -x)$; this implies that if $Q(x; c)$ is a 
solution of
(\ref{BEml}), so is $Q(-x, ;-c)$.

We now determine the qualitative
nature of a solution $Q(x)$ for the symmetric 
 polynomial $T$-$Q$ equation. 
\begin{lemma} \label{lem:symNo} 
For a
symmetric $T$-$Q$ polynomial relation
$(\ref{BEml})$, there is no non-trivial
solution $Q(x)$ of  degree
$d=N-2$ with the zero-multiplicity at $x=0$ equal
to
$N-2m$.
\end{lemma}
{\it Proof.} Otherwise, one has $m \geq 1$
and  
\be
Q(x) = x^{N-2m}\widetilde{Q}(x) \ , \ \ \ {\rm
where } \ \ 
\widetilde{Q}(0) \neq 0 \ , \ {\rm
deg} \widetilde{Q} = 2m-2 \ .
\ele(QQt)
Write $
\widetilde{Q} (x) = \sum_{j=0}^{2m-2}
\widetilde{\alpha}_j x^j$. 
Then $\widetilde{Q}(x)$ satisfies the relation
\begin{eqnarray}
 \Lambda_{m,l} (x) 
\widetilde{Q} (x) = q^m \triangle( x (-
q)^{\frac{-1}{2}}) \widetilde{Q}(xq^{-1})+ q^{-m}
\triangle(x (- q)^{\frac{1}{2}}) 
\widetilde{Q} (xq) \ , \label{Qtil}
\end{eqnarray}
or equivalently, the coefficients
$\widetilde{\alpha}_j$s of
$\widetilde{Q} (x)$ satisfy the following system
of equations, 
\be
\widetilde{\nu}_j \widetilde{\alpha}_j +
\widetilde{ v}_j 
\widetilde{\alpha}_{j-1} +
(\widetilde{\delta}_j - \lambda)
\widetilde{\alpha}_{j-2} +\widetilde{u}_j
\widetilde{\alpha}_{j-3} +  \widetilde{\mu}_j
\widetilde{\alpha}_{j-4} = 0 \ , \ \  1 \leq j
\leq 2m+1 \ ,
\ele(Qtide)
where $\widetilde{\nu}_j,
\widetilde{v}_j,
\widetilde{\delta}_j, \widetilde{u}_j,
\widetilde{\mu}_j$ are expressed by the
similar forms as in (\req(coefsy)) by
changing $m$ to $-m$ in the corresponding term.
By
$\widetilde{\nu}_1 \neq 0$, we have $m \geq 2$. By
the equalities,
$$
\widetilde{\nu}_j =
\widetilde{\mu}_{2m+2-j},  \ \ \widetilde{v}_j =
\widetilde{u}_{2m+2-j},  \ \ \widetilde{\delta}_j
=
\widetilde{\delta}_{2m+2-j} \ ,
$$
$\widetilde{Q}^\dagger(x)$ also satisfies
the equation (\ref{Qtil}). In general, for a
polynomial $\widetilde{Q}(x)$ of degree
$\tilde{d}$ satisfies (\ref{Qtil}), $\tilde{d}
\equiv 2m-2, N-2 \pmod{N}$, and the minimal
possible
$\tilde{d}$ is $2m-2$. Hence the dimension of the
solution space of $\widetilde{Q}(x)$ with degree
$\leq 2m-2$ is equal to one. For 
$\widetilde{Q}(x)$ in (\req(QQt)),
$\widetilde{Q}^\dagger(x)$ is a
scale-multiple of $\widetilde{Q}(x)$, which
implies $
\widetilde{Q}^\dagger(x)= \pm
\widetilde{Q}(x)$. Therefore, $\widetilde{Q}(x)$
is determined by the coefficients
$\widetilde{\alpha}_j$ for $ 0
\leq j \leq m-1$, which involve only those
equations in (\req(Qtide)) with 
$1 \leq j \leq m+1$, subject to one
of the following two conditions:
$\widetilde{\alpha}_j=
\widetilde{\alpha}_{2m-2-j}$ for all $j$, or
$\widetilde{\alpha}_j= -
\widetilde{\alpha}_{2m-2-j}$ for all $j$.  
On the other hand, the relation (\ref{Qtil}) is
the same when we substitute $(c, x)$ by $(-c,
-x)$, hence $\widetilde{Q}(-x, -c)=
\widetilde{Q}(x, c)$. Hence we may
assume the coefficients
$\widetilde{\alpha}_j= \widetilde{\alpha}_j(c)$
satisfy the following properties:
\be
\widetilde{\alpha}_0(c) =
1 \ , \ \ \ \ \widetilde{\alpha}_j(-c) =
(-1)^j\widetilde{\alpha}_j(c) , \ \ {\rm for \
all} \ j \ .
\ele(cofeod)
For a
solution $\{ \lambda , 
\widetilde{\alpha}_j \}$ of (\req(Qtide)) for
a generic $c$,
$\lambda$ is a solution of the eigenvalues problem
arisen from those relations for  $2 \leq
j \leq m$. Hence
$\lambda=
\lambda (c)$, an algebraic function of $c$
so that the limit of 
$\frac{\lambda(c)}{c^2}$, denoted by
$\lambda_\infty$, exists as $c
\rightarrow
\infty$. For $1 \leq j
\leq m-1$, by $\widetilde{\nu}_j
\neq 0$  one can conclude  
$\widetilde{\alpha}_j(c) = O( c^j)$  as $c
\rightarrow \infty$. Denote 
$$
{\sf a}_j =  \lim_{c \rightarrow \infty}
\frac{\widetilde{\alpha}_j(c)}{c^j} , \ \ \ \ \ 0
\leq j \leq m-1 \ ,
$$
and 
$$
\widetilde{v}_k^\prime =  4{\rm
i}(q^{-m+k-\frac{1}{2}}- q^{m-k+\frac{1}{2}}) ,
\ \  
\widetilde{u}_k^\prime = - 4{\rm
i}(q^{-m+k-\frac{3}{2}}- q^{m-k+\frac{3}{2}}) ,\ \
\widetilde{\delta}_k^\prime =  -4(q^{-m+k-1}+
q^{m-k+1})  
$$
for $1 \leq k \leq m+1 $.
By multiplying $c^{-j}$ on (\req(Qtide)), and then
taking the $c$-infinity limit for
$1
\leq j \leq m+1$, one obtains the following matrix
relation on $\lambda_\infty$ and ${\sf a}_j$s,
\begin{eqnarray}
 \left( \begin{array}{ccccccc}
\widetilde{\delta}^\prime_{m+1}-\lambda_\infty
&0&0& \cdots&
&  &0 \\
\widetilde{v}^\prime_{m}&
\widetilde{\delta}^\prime_{m}-\lambda_\infty
&0&0&\ddots   & & \vdots
\\ 
\widetilde{\nu}_{m-1}&\widetilde{
v}^\prime_{m-1}&
\widetilde{\delta}^\prime_{m-1}-\lambda_\infty&
0&\ddots  &
\ddots &\vdots \\
 0& \ddots&\ddots &\ddots &  & \ddots
&\vdots \\
 \ddots & \ddots & \ddots && \ddots & 
\ddots &0\\
\vdots& \ddots   &\widetilde{\nu}_4
&\widetilde{v}^\prime_4 &
\widetilde{\delta}^\prime_4-\lambda_\infty &0 &0\\
\vdots  &\ddots  &\ddots &\widetilde{\nu}_3
&\widetilde{v}^\prime_3 &
\widetilde{\delta}^\prime_3-\lambda_\infty &0\\ 
0&  \cdots &  \cdots & 0
&\widetilde{\nu}_2& \widetilde{v}^\prime_2
&
\widetilde{\delta}^\prime_2-\lambda_\infty
\\ 0&  \cdots&   & \cdots
&0&\widetilde{\nu}_1&
\widetilde{v}^\prime_1
\end{array} 
\right) 
 \left( \begin{array}{c}
{\sf a}_{m-1}\\
{\sf a}_{m-2} \\
\vdots \\
\vdots \\
\vdots \\
\vdots \\
\vdots\\
{\sf a}_0
\end{array} 
\right) = \vec{0}  \ . \label{Mat0}
\end{eqnarray}
Note that ${\sf
a}_0 = 1$, and $
\widetilde{\delta}^\prime_j
 \neq \widetilde{\delta}^\prime_k $ for $  2
\leq j \neq  k \leq m+1$. The square matrix by
deleting the last row in (\ref{Mat0})
becomes the eigenvalue problem with $m$ distinct
eigenvalues, hence 
$$
\lambda_\infty =
\widetilde{\delta}^\prime_l \ , \ {\rm for \ some
} \ 2 \leq l \leq m+1 . 
$$
This implies ${\sf a}_k= 0 $ for
$l-2 < k \leq m-1$, and ${\sf a}_{l-2} \neq 0$.
In (\ref{Mat0}), the row containing
$\widetilde{\delta}^\prime_{l-1}-\lambda_\infty$
 gives the following
relation of ${\sf a}_l$ and ${\sf a}_{l-1}$:
\be
\widetilde{v}^\prime_{l-1} {\sf a}_{l-2} + 
(\widetilde{\delta}^\prime_{l-1}-
\widetilde{\delta}^\prime_l
){\sf a}_{l-3} = 0 \ .
\ele(lrel)
One would expect the right lower square matrix of
size $l-1$ with  $\lambda_\infty =
\widetilde{\delta}^\prime_l$ has non-zero
determinant. This statement is valid for
a small number
$l$ by direct computation, hence 
it leads to a contradiction. However, it
is a  difficult task to obtain a  mathematical
proof of such a statement for a general $l$. For
our purpose, we are going to provide another way
to justify the conclusion of the lemma using
(\req(Qtide)).   First we consider the case
$l= m+1$, then ${\sf a}_{m-1} \neq 0$. This 
implies $\widetilde{Q}^\dagger (x) =
\widetilde{Q}(x) $.  The relation (\req(lrel))
and the
$(m+1)$th equation  in (\req(Qtide)) become  
\bea(l)
\widetilde{v}^\prime_{m} {\sf a}_{m-1} + 
(\widetilde{\delta}^\prime_{m}-
\widetilde{\delta}^\prime_{m+1}
){\sf a}_{m-2} = 0 \ , \\
(\widetilde{\delta}_{m+1} - \lambda (c))
\widetilde{\alpha}_{m-1} +2\widetilde{u}_{m+1}
\widetilde{\alpha}_{m-2} + 2 \widetilde{\mu}_{m+1}
\widetilde{\alpha}_{m-3} = 0 \ .
\elea(am1)
By the property (\req(SyDef)) on $\lambda$, the
$c$-infinity limit of 
$c^{-(m-1)}$-multiple of the second relation  in
(\req(am1)) gives rise to the equality,
$$
-4{\sf a}_{m-1} +2\widetilde{u}^\prime_{m+1}
{\sf a}_{m-2}  = 0 \ ,
$$
which is incompatible with the first relation of
(\req(am1)). 
Hence, we may assume $2 \leq l \leq m$. By
(\req(cofeod)) and ${\sf a}_k=0$  for $l-2 <
k$, one has 
$$
\widetilde{\alpha}_k (c) = {\sf a}_k^{\prime}
c^{k-2} + {\rm lower \ order \ term} \ ,
\ \ {\rm as} \  c \rightarrow \infty \ , \ \ \
\ \ \ l-1 \leq k \leq m-1. 
$$
By $\lambda_\infty = \widetilde{\delta}_l^\prime$,
 one can take the $c$-infinity
limit of $c^{-(l-2)}$-multiple of the $l$th
relation in (\req(Qtide)), which yields  the
following identities: 
\bea(ll)
\widetilde{\nu}_l {\sf a}_l^{\prime} +
\widetilde{ v}^\prime_l 
{\sf a}_{l-1}^{\prime} +
\frac{\widetilde{\delta}^\prime_l }{2}
{\sf a}_{l-2} +\widetilde{u}^\prime_l
{\sf a}_{l-3} = 0 \ , & {\rm when} \  l \leq m-1 ;
\\
\widetilde{\nu}_m {\sf a}_m +
\widetilde{ v}^\prime_m 
{\sf a}_{m-1}^{\prime}
+ \frac{\widetilde{\delta}^\prime_m }{2}{\sf
a}_{m-2} +\widetilde{u}^\prime_m {\sf a}_{m-3} =
0 \ , & {\rm when} \ l = m .
\elea(linf)
Similarly  the $c$-infinity
limit of 
$c^{-(l-1)}$-multiple of the $(l+1)$relation
gives rise to the following relations:
\bea(ll)
\widetilde{\nu}_{l+1} {\sf a}^\prime_{l+1} +
\widetilde{ v}^\prime_{l+1} 
{\sf a}^\prime_{l} +
(\widetilde{\delta}^\prime_{l+1} -
\widetilde{\delta}^\prime_l)
{\sf a}^\prime_{l-1}
+\widetilde{u}^\prime_{l+1}
{\sf a}_{l-2}  = 0 \  , & {\rm when} \ l \leq m-2
; \\
\widetilde{ v}^\prime_{m}
{\sf a}^\prime_{m-1} +
(\widetilde{\delta}^\prime_{m} -
\widetilde{\delta}^\prime_{m-1})
{\sf a}^\prime_{m-2}
+\widetilde{u}^\prime_{m}
{\sf a}_{m-3}  = 0 \  , & {\rm when} \ l = m-1 ;
\\ (\widetilde{\delta}^\prime_{m+1} -
\widetilde{\delta}^\prime_m)
{\sf a}^\prime_{m-1}
+\widetilde{u}^\prime_{m+1}
({\sf a}_{m-2}+{\sf a}_m)  = 0 \  , & {\rm when}
\ l =m . \\
\elea(l-inf)
For $l=m$, by using ${\sf a}_m = \pm {\sf
a}_{m-2}$,  (\req(lrel)) (\req(l-inf)) and the
last relation in (\req(linf)) will lead
to a contradiction. For $l \leq m-2$, we continue
the same procedure on the $c$-infinity
limit of $c^{-s}$-multiple of the $(s+2)$th
relation for $s \geq l$, then obtain 
$$
\widetilde{\nu}_{s+2} {\sf a}^\prime_{s+2} +
\widetilde{ v}^\prime_{s+2} 
{\sf a}^\prime_{s+1} +
(\widetilde{\delta}^\prime_{s+2} -
\widetilde{\delta}^\prime_s)
{\sf a}^\prime_{s}  = 0 \ .
$$
Hence one has the following relations for ${\sf
a}_k^\prime$s,
\begin{eqnarray*}
 \left( \begin{array}{cccccc}
\widetilde{\delta}^\prime_{m+1}-
\widetilde{\delta}^\prime_l
&0&0& \cdots&
  &0 \\
\widetilde{v}^\prime_{m}&
\widetilde{\delta}^\prime_{m}-
\widetilde{\delta}^\prime_l
&0&0&\ddots   & \vdots
\\ 
\widetilde{\nu}_{m-1}&\widetilde{
v}^\prime_{m-1}&
\widetilde{\delta}^\prime_{m-1}-
\widetilde{\delta}^\prime_l&
0&\ddots   &\vdots \\
 0& \ddots&\ddots &\ddots &   \ddots
&\vdots \\
 \ddots & \ddots & \ddots && \ddots  &0\\
\vdots& \ddots   &\widetilde{\nu}_{l+3}
&\widetilde{v}^\prime_{l+3} &
\widetilde{\delta}^\prime_{l+3}-
\widetilde{\delta}^\prime_l
&0 \\
\vdots  &\ddots  &\ddots &\widetilde{\nu}_{l+2}
&\widetilde{v}^\prime_{l+2} &
\widetilde{\delta}^\prime_{l+2}-
\widetilde{\delta}^\prime_l
\end{array} 
\right) 
 \left( \begin{array}{c}
{\sf a}^\prime_{m-1}\\
{\sf a}^\prime_{m-2} \\
\vdots \\
\vdots \\
\vdots \\
\vdots \\
\vdots\\
{\sf a}^\prime_{l}
\end{array} 
\right) = \vec{0}  ,
\end{eqnarray*}
which implies ${\sf a}^\prime_s = 0$ for $s \geq
l$. The relations (\req(linf)) (\req(l-inf))
become
$$
\widetilde{ v}^\prime_l 
{\sf a}_{l-1}^{\prime} +
\frac{\widetilde{\delta}^\prime_l }{2}
{\sf a}_{l-2} +\widetilde{u}^\prime_l
{\sf a}_{l-3} = 0 \ , \ \ 
(\widetilde{\delta}^\prime_{l+1} -
\widetilde{\delta}^\prime_l)
{\sf a}^\prime_{l-1}
+\widetilde{u}^\prime_{l+1}
{\sf a}_{l-2}  = 0 \ , 
$$
together with (\req(lrel)), this provides a
contradiction to ${\sf a}_{l-2} \neq 0 $. 
$\Box$ \par \vspace{0.2in} \noindent
For a symmetric (\ref{BEml}) equation, the
eigen-polynomial $Q(x)$ has the following
property:
\begin{theorem} \label{thm:syQ} 
Assume there exists a non-trivial polynomial
solution for a symmetric $T$-$Q$ polynomial
relation
$(\ref{BEml})$ with a given reciprocal polynomial
$\Lambda_{m,l}(x)$. Then the equation has
one-dimensional solution space,   generated by a
monic polynomial $Q(x)$ of degree
$2N-2-2m$ with $Q(0) \neq 0 $ and $Q^\dagger(x)=
\pm Q(x)$.
\end{theorem}
{\it Proof.} Let $Q(x)$ be a non-trivial
polynomial solution with the degree $d$. By Lemma
\ref{lem:d0}, we have
$d = N-2 , N-2m-2, 2N-2m-2$. First we are going to
show that
$d= 2N-2m-2$. Otherwise, $d$ is one of two odd
integers,
$N-2m-2$ or
$ N-2$. By Lemmas \ref{lem:d0} and
\ref{lem:symNo}, we may assume
$Q(0) \neq 0
$ with $\alpha_0=1$. The coefficients 
$\alpha_j$s of
$Q(x)$ satisfy the relation
(\req(eqDL4)). When
$d=N-2m-2$, we have 
$ \nu_j \neq 0 $ for $1 \leq j \leq d$. This
implies the $x$-coefficients $\alpha_j$ of 
$Q(x)(= Q(x;c))$ are polynomials of
$c$ and $\lambda=\lambda(c)$, hence $\alpha_j =
\alpha_j(c)$. As $c$ tends to $0$, the
coefficients $\alpha_j=\alpha_j(0)$ of 
$Q(x;0)$  satisfy the corresponding relation
(\req(eqDL4)):
$$
\nu_j \alpha_j + 
(\delta_j - \lambda) \alpha_{j-2} +  \mu_j
\alpha_{j-4} = 0 \ \ \ , \ 1 \leq j \leq d+3 \ ,
$$
with $\alpha_1=0$. Hence $\alpha_j=0$ for odd
$j$, and the polynomial $Q(x;0)$ has an even
degree
$\leq N-2m-2$, which impossible by Lemma
\ref{lem:d0}. It remains the case when 
$d= N-2$ with $Q(0;c) \neq 0$. Now the
dimension of the $Q(x)$-solution space of 
(\ref{BEml})  with ${\rm deg} \ Q(x)
\leq N-2$  is equal to one. As 
$Q(-x, ;-c)$ is also a solution of the symmetric
$T$-$Q$ relation,  we have 
$Q(-x, ;-c)= Q(x ; c)$, equivalently ,
$\alpha_j (-c)= (-1)^j
\alpha_j(c)$ for all $j$. Therefore $Q(x ; 0)$ is
again a polynomial in $x$ with an even degree
$\leq N-2$, which contradicts to Lemma
\ref{lem:d0} for $c=0$. Hence we have shown that
any solution
$Q(x)$ with the eigenvalue $\Lambda_{m,l}(x)$
must have the degree $d = 2N-2-2m$, which implies
the $Q(x)$-solution space is of dimension 
one.   By (\req(c*)), the polynomials, 
$\Lambda_{m,l}(x)$ and $ Q(x)$, satisfy  the
conditions of Proposition
\ref{prop:syTQ},
hence
$Q^\dagger(x)$ is a also a solution of 
(\ref{BEml}). Therefore $ Q^\dagger(x) = \gamma
Q(x)$ for some non-zero constant $\gamma$, which
implies $\gamma^2=1$ and $Q(0) \neq 0$. Then the
conclusion follows immediately.
$\Box$ \par \vspace{0.2in} \noindent
{\bf Remark.} For a polynomial $Q(x)$ in
the above proposition, the roots of
$Q(x)$ are all non-zero; furthermore if $x_k$ is a
root, so is  $x_k^{-1}$. Hence the collection of
all roots $x_k$ (counting multiplicity) is
the same as that of
$x_k^{-1}$s. The criterion for 
$Q^\dagger(x) = - Q(x)$ holds if and only if
$Q(x)$ has the root $x=1$ with a positive odd
multiplicity.
$\Box$ \par \vspace{0.2in} \noindent

\section{Solutions of 
Discrete Quantum Pendulum and Sine-Gordon
Model in the Rational Degenerated case}   In this
section we are going to derive the complete
solution of symmetric
$T$-$Q$ polynomial relation (\ref{BEml}); hence
we now only consider the sectors $(m, l)= (m,
0), (m, N-2m)$. By Theorem
\ref{thm:syQ}, we may assume  
$$ 
d=2N-2-2m \ , \ \ Q^\dagger(x) = \pm Q(x) . 
$$
Hence the coefficients in 
(\req(coefsy)) possesses the following symmetric
relations:  
$$
\nu_{d+4-j}= \mu_j , \ \  v_{d+4-j}= u_j , \ \ 
\delta_{d+4-j} = \delta_j \ .
$$
The system (\req(eqDL4)) is equivalent to the
eigenvalue problem (\req(Mform)) together with
one more constraint:
\be
\nu_1 \alpha_1 + v_1 \alpha_0  = 0 \ , 
\ele(symetra)
and the $\alpha_j$s satisfy either one
of the following conditions :
\begin{eqnarray}
\alpha_i = \alpha_{d-i} \ & {\rm for} \ 0 \leq i
\leq d ,\  & i.e., \ Q^\dagger(x) = Q(x) ;
\label{QQd}
\\
\alpha_i = -\alpha_{d-i} \ &{\rm for} \ 0  \leq i
\leq d , \ & i.e., \ Q^\dagger(x) = -Q(x) .
\label{QQd-}
\end{eqnarray}
Note that the polynomial $Q(x)$ is 
determined only by  the first half of its
coefficients, i.e., $\alpha_0,
\ldots, \alpha_{\frac{d}{2}}$. In the
case  (\ref{QQd-}), one has
$\alpha_{\frac{d}{2}}=0$.
Furthermore, through  the transformations: 
$$
\begin{array}{l}
\nu_j, v_j, \delta_j, u_j, \mu_j \mapsto
\mu_{j'}, u_{j'}, \delta_{j'}, v_{j'}, \nu_{j'} \
, \ \ \ {\rm where } \ \ j':= d+4-j \ \ ;
\\
\alpha_i \mapsto \alpha_{d-i}  \ (0 \leq i \leq
d ) \  \ \  {\rm or} \ \ \ \ \alpha_i \mapsto
-\alpha_{d-i} \ \ ( 0 \leq i \leq d ) \ , 
\end{array}
$$
the equations for $j \geq
\frac{d}{2}+3$ in (\req(eqDL4))  follow from 
those  for $j \leq \frac{d}{2}+2$. So we need
only to consider the relations for $1 \leq j \leq
\frac{d}{2}+2$ in (\req(eqDL4)), which are
regarded as equations of
$\lambda$ and $\alpha_k$ for $ 0 \leq k \leq
\frac{d}{2}$. Note that
the $(\frac{d}{2}+2)$th equation in
(\req(eqDL4)) has the form :
\be
(\delta_{\frac{d}{2}+2}-\lambda)
\alpha_{\frac{d}{2}} + 
u_{\frac{d}{2}+2}(\alpha_{\frac{d}{2}-1}
+\alpha_{\frac{d}{2}+1})
+ \mu_{\frac{d}{2}+2} ( \alpha_{\frac{d}{2}-2}
+\alpha_{\frac{d}{2}+2}) = 0 \ \ ,
\ele(QQdc)
which is a trivial relation in the case
(\ref{QQd-}).

By (\req(DQPSG)), 
the rational degenerated case of  discrete
quantum pendulum and discrete
sine-Gordon  corresponds to $C=1$,
i.e. the sectors with $l=0$ in the
symmetric (\ref{BEml}) relation; in particular,
by (\req(DQP)) and (\req(deg4)),  the discrete
quantum pendulum is given by 
$D=C=1$, i.e. $(m,
l)=(0, 0)$ . 
\begin{theorem} \label{thm:SMTQ}
For the symmetric
$T$-$Q$ polynomial equation
$(\ref{BEml})$,  there
are
$N$ distinct eigenvalues
$\lambda$, each of which  has  one-dimensional
eigenspace generated by a monic 
eigen-polynomial
$Q(x)$ of degree $d=4M-2m$ with $Q(0) \neq 0$ and
$Q^\dagger(x) = \pm Q(x)$. Furthermore, 
there are
$(M+1)$ eigen-polynomials $Q(x)$ of the
type
$Q^\dagger(x) = Q(x)$, and the rest $M$ ones
are of the type $Q^\dagger(x) = - Q(x)$. In
particular, the Baxter's $T$-$Q$
polynomial relation of ${\rm SG}$
model are those for the  sectors
$(m, l)=(m, 0)$, and the discrete quantum pendulum
is the one for $(m, l)=(0,0)$.
\end{theorem}
{\it Proof.} The relation (\req(symetra)) is a
non-trivial constraint for $0 \leq m \leq
M-1$ by $\nu_1 \neq 0$; while for $m=M$, both
$\nu_1$ and
$v_1$ are zeros, hence (\req(symetra)) is a
redundant one. In this proof, we
shall first consider the case with $m=0$, then $1
\leq m \leq M-1$, and finally on $m=M$.

(I)  $m=0$, i.e.
$(m, l)=(0,0)$, which is the rational degenerated
 case of 
 discrete quantum
pendulum. We have $d=4M$. Consider
the relations for $1 \leq j \leq \frac{d}{2}+2$ 
in the system (\req(eqDL4)) as equations of 
$\lambda, \alpha_0 ,...,
\alpha_{\frac{d}{2}}$. By
$\nu_{\frac{d}{2}+1}=0$, the problem is formulated
in the following matrix form:
\begin{eqnarray}
 \left( \begin{array}{cccccccc}
v_{\frac{d}{2}+1}&
\delta_{\frac{d}{2}+1}-\lambda
&u_{\frac{d}{2}+1}&\mu_{\frac{d}{2}+1}&0& \cdots
&  &0 \\
\nu_{\frac{d}{2}}&v_\frac{d}{2}&
\delta_\frac{d}{2}-\lambda
&u_\frac{d}{2}&\mu_\frac{d}{2}&\ddots   & & \vdots
\\ 0& \nu_{\frac{d}{2}-1}&v_{\frac{d}{2}-1}&
\delta_{\frac{d}{2}-1}-\lambda&\ddots &\ddots  &
\ddots &\vdots \\
 0& \ddots& \ddots&\ddots &\ddots &  & \ddots
&\vdots \\
\vdots& \ddots & \ddots & \ddots && \ddots & 
\ddots &0\\
\vdots& \ddots& \ddots   &\nu_4 &v_4 &
\delta_4-\lambda &u_4&\mu_4\\
\vdots& \ddots  &\ddots  &\ddots &\nu_3 &v_3 &
\delta_3-\lambda &u_3\\ 
0&  \cdots & & \cdots & 0
&\nu_2& v_2 & \delta_2-\lambda
\\ 0&  \cdots&  \cdots & & \cdots &0&\nu_1& v_1
\end{array} 
\right) 
 \left( \begin{array}{c}
\alpha_{\frac{d}{2}}\\
\alpha_{\frac{d}{2}-1} \\
\vdots \\
\vdots \\
\vdots \\
\vdots \\
\vdots\\
\alpha_0
\end{array} 
\right) = \vec{0}  \ \label{Mat00}
\end{eqnarray}
together with the constraint (\req(QQdc)), which
is now given by
\be
 (q +q^{-1}-2)(\alpha_{\frac{d}{2}+2}
+\alpha_{\frac{d}{2}-2})+
 4c{\rm i}(q^{\frac{1}{2}}-q^{\frac{-1}{2}}) 
(\alpha_{\frac{d}{2}+1} +\alpha_{\frac{d}{2}-1}) -
(8c^2+4  +
\lambda) \alpha_{\frac{d}{2}} = 0 \ .
\ele(OOex)
Note that the square matrix of size $N \ (=
\frac{d}{2}+1)$ in 
(\ref{Mat00}) satisfies the condition of Lemma
\ref{lem:det}. Hence, by $\nu_j \neq 0$ for
$1 \leq j \leq
\frac{d}{2}$, the system (\ref{Mat00}) has the
one-dimensional eigenspace for any given 
$c$ and $\lambda$, generated by a basis element
$(\alpha_k)_{0\leq k \leq \frac{d}{2}}$ with
$\alpha_0=1$. In fact, for $1
\leq k \leq  \frac{d}{2}$, 
$\alpha_k$ can be expressed by a
polynomial of $\lambda$ and $ c$, 
regarded  as a polynomial in $\lambda$ with
coefficients in
$\CZ[c]$, and denoted by
$\alpha_k = p_k(\lambda)$.  The
$\lambda$-degree of $\alpha_k$ is given by
${\rm deg} \ p_k  (\lambda) =
[\frac{k}{2}]$. 
In  the case (\ref{QQd}), the relation
(\req(OOex)) becomes:
$$
 2(q +q^{-1}-2) p_{\frac{d}{2}-2}(\lambda) +
 8c{\rm i}(q^{\frac{1}{2}}-q^{\frac{-1}{2}})
p_{\frac{d}{2}-1}(\lambda) - (8c^2+4  +
\lambda) p_\frac{d}{2}(\lambda)  = 0 \ , 
$$
which defines 
$\lambda$ as an algebraic function of $c$.  
As the $\lambda$-degree of the above relation is
equal to 
$M+1$, there are $(M+1)$
$\lambda$-values for a generic $c$. 
In the case (\ref{QQd-}),  (\req(OOex)) is
a trivial relation. The relation  (\ref{Mat00})
becomes: 
$$
\begin{array}{l}
 \left( \begin{array}{ccccccc}
 \delta_{\frac{d}{2}+1}-\lambda
&u_{\frac{d}{2}+1}&\mu_{\frac{d}{2}+1}&0 &
\cdots&  &0
\\
v_\frac{d}{2}& \delta_\frac{d}{2}-\lambda
&u_\frac{d}{2}&\mu_\frac{d}{2}&0 & \cdots  &0
\\
\nu_{\frac{d}{2}-1}&v{\frac{d}{2}-1}&\delta_{\frac{d}{2}-1}
-\lambda
&u_{\frac{d}{2}-1}&\mu_{\frac{d}{2}-1}&\ddots   &0
 \\
0& \ddots& \ddots&\ddots &\ddots  & \ddots
&\vdots \\
\vdots& \ddots & \ddots & \ddots & \ddots & 
\ddots &0\\
\vdots& \ddots   &\nu_4 &v_4 &
\delta_4-\lambda &u_4&\mu_4\\
\vdots& \ddots  & \ddots &\nu_3 &v_3 &
\delta_3-\lambda &u_3\\ 
0&  \cdots & & 0
&\nu_2& v_2 & \delta_2-\lambda
\\ 0&  \cdots & & 0 &0&\nu_1& v_1
\end{array} 
\right) 
 \left( \begin{array}{c}
\alpha_{\frac{d}{2}-1} \\
\vdots \\
\vdots \\
\vdots \\
\vdots\\
\alpha_0
\end{array} 
\right) = \vec{0} \ .
\end{array}
$$
Solutions of the above equation can be obtained
from the system (\ref{Mat00}) alone, and then 
imposing the following constraint on $\lambda$,
$$
\alpha_{\frac{d}{2}} = p_{\frac{d}{2}}(\lambda) =
0
\ .
$$
As the
$\lambda$-degree of $p_k  $ is equal
to $M$,
 the above equation
gives rise to $M$ eigenvalues of $\lambda$, with 
the corresponding  eigenvector having the
components
$\alpha_k = p_k(\lambda), 1 \leq k \leq
\frac{d}{2}-1$. By
Theorem \ref{thm:syQ}, the $Q(x)$-eigenspaces 
are all one-dimensional, hence the conclusion
follows immediately.

(II) $1 \leq m \leq M-1$. We have $l=0, N-2m $,
and $d= 2N-2-2m$. By (\req(coefsy)), for $
1 \leq j \leq \frac{d}{2}+1$ one has
\begin{eqnarray}
\nu_j=0 & \Leftrightarrow & j=
{\bf n}:=  N-2m \ .
\label{bfn}
\end{eqnarray}
In the case (\ref{QQd}), the relations
for
$1 \leq j \leq \frac{d}{2}+2$ in the system
(\req(eqDL4)), considered as equations of
$\lambda$ and $\alpha_k \ (0 \leq k \leq
\frac{d}{2})$, can be formulated in the
following form,
\be
 \left( \begin{array}{cc}
S&T \\
0 &U
\end{array} 
\right)  
 \left( \begin{array}{c}
\widetilde{\psi} \\
\psi
\end{array} 
\right) = \vec{0} \ , \ \ \ \ \ \
\psi:= \left( \begin{array}{c}
\alpha_{{\bf n}-1} \\
\vdots\\
\vdots\\
\alpha_1 \\
\alpha_0
\end{array} 
\right) \ , \ \ \
\widetilde{\psi}:= 
\left( \begin{array}{c}
\alpha_{\frac{d}{2}} \\
\vdots \\
\vdots \\
\alpha_{{\bf n}+1} \\
\alpha_{\bf n} 
\end{array} 
\right) \ ,
\ele(STU)
and together with the  constraint
(\req(QQdc)). Here $S,  U $ are  square
matrices of the size $(\frac{d}{2}-{\bf n}+1)$,
${\bf n}$ respectively,  $T$ is  the 
$(\frac{d}{2}-{\bf n}+1) \times {\bf n}$ matrix,
with the following expressions:
$$
\begin{array}{l}
S= \left( \begin{array}{ccccccc}
v_{\frac{d}{2}+1} &\nu_{\frac{d}{2}+1}+
\delta_{\frac{d}{2}+1}-\lambda
&u_{\frac{d}{2}+1}&\mu_{\frac{d}{2}+1}&0 &
\cdots &0
\\
\nu_{\frac{d}{2}} &v_{\frac{d}{2}}&
\delta_{\frac{d}{2}}-\lambda
&u_{\frac{d}{2}}&\mu_{\frac{d}{2}}&0 & \vdots
\\
0& \ddots & & & & &
\\
\vdots &0 &\ddots & & & & \vdots 
\\
\vdots &\cdots &\cdots & & & &  
\\
\vdots &\cdots &0&\nu_{{\bf n}+3}&v_{{\bf n}+3}&
\delta_{{\bf n}+3}-\lambda &u_{{\bf n}+3}
\\
0&\cdots& \cdots &0&\nu_{{\bf n}+2}&v_{{\bf n}+2}&
\delta_{{\bf n}+2}-\lambda
 \\ 
0& \cdots & \cdots & & 0 &\nu_{{\bf n}+1}&v_{{\bf
n}+1} 
\end{array} 
\right) \ , \\
U= \left( \begin{array}{ccccccc}
v_{\bf n}&
\delta_{\bf n}-\lambda
&u_{\bf n}&\mu_{\bf n}&0&\cdots   &0
\\
\nu_{{\bf n}-1}&v_{{\bf n}-1}&
\delta_{{\bf n}-1}-\lambda
&u_{{\bf n}-1}&\mu_{{\bf n}-1}&0&\vdots   
\\ 
0 & \ddots&\ddots &\ddots  &
\ddots &&\vdots \\
\vdots & \ddots&\ddots &\ddots  &
\ddots & \ddots &0 \\
\vdots& \ddots & \ddots & \ddots & \ddots
& 
\ddots &\mu_4\\
&\cdots  & 0 &\nu_3 &v_3 &
\delta_3-\lambda &u_3\\
\vdots& \cdots & & 0
&\nu_2& v_2 & \delta_2-\lambda
\\ 
0& \cdots & & 0 &0&\nu_1& v_1
\end{array} 
\right) \ , \\ 
T=  \left( \begin{array}{cccccc}
0 & \cdots & \cdots & &\cdots & 0
\\
\vdots & & \cdots & & &\vdots
\\
 0& \cdots&\cdots &  &
 &\vdots \\
\mu_{{\bf n}+3}&0&\cdots &  &&\vdots
\\
u_{{\bf n}+2}&0 &\cdots & & & 0
\\
\delta_{{\bf n}+1}-\lambda
&u_{{\bf n}+1}&\mu_{{\bf n}+1}&0  & \cdots  &0
\end{array} 
\right) \ .
\end{array}
$$
Note that there is the term $\nu_{\frac{d}{2}}$
in the second entry of the first
row of $S$; while $\mu_{{\bf n}+2}=0$ in the
matrix $T$. By (\req(coefsy)), the matrix $U$
satisfies the condition of Lemma \ref{lem:det}
for any $\lambda$. Hence, by (\ref{bfn}), the
system $U \psi = 0 $ inside the
relation (\req(STU))
 has the
one-dimensional solution space generated by a
vector
$\psi$ with  
$$
\alpha_0 = 1 , \ \ \  \alpha_k = p_k ( \lambda)
\in \CZ[c][ \lambda]  ,  \ \ \ {\rm where} \ \ \ 
{\rm deg} \ p_k (\lambda) = [
\frac{k}{2} ]  \ \ \  {\rm for} \ \ k <
{\bf n} .
$$
Using the above vector $\psi$, we consider the
system 
$$
S \widetilde{\psi} = - T \psi \ .
$$
By (\ref{bfn}), one can first solve $\alpha_k \ ( 
k > {\bf n}) $ in terms of
$\alpha_{\bf n}, \lambda,
c$ with the form : 
$$
\begin{array}{l}
\alpha_k = r_k(\lambda) \alpha_{\bf n} + q_k
(\lambda) , \ \ {\rm where} \ \ r_k(\lambda), 
q_k
(\lambda) \in \CZ[c][\lambda] \ , 
\ \ \ 
{\rm deg} \ r_k (\lambda) + \frac{{\bf
n}-1}{2} ={\rm deg} \ q_k (\lambda) =
[\frac{k}{2}]. 
\end{array}
$$
Furthermore,  $\alpha_{\bf n}$ satisfies the
following relation :
$$
\begin{array}{l}
0= v_{\frac{d}{2}+1} \alpha_{\frac{d}{2}} +
( \nu_{\frac{d}{2}+1}+\delta_{\frac{d}{2}+1}
-\lambda)\alpha_{\frac{d}{2}-1} +
u_{\frac{d}{2}+1}\alpha_{\frac{d}{2}-2}+
\mu_{\frac{d}{2}+1}\alpha_{\frac{d}{2}-3} = r(
\lambda) \alpha_{\bf n} + q (\lambda ) \ , 
\end{array}
$$
where $r(\lambda), q(\lambda) \in
\CZ[c][\lambda]$ with
$ {\rm deg} \ r (\lambda) + \frac{{\bf
n}-1}{2} = {\rm deg} \ q(\lambda) =
[\frac{d+2}{4}]$. 
 Hence 
$$
\alpha_k =
\frac{P_k(\lambda)}{r(\lambda)}
\ , \ \ \ \ 
P_k(\lambda) := r(\lambda)q_k(\lambda)
- r_k(\lambda)q(\lambda)  \ .
$$
By multipling the
$\alpha_k$ by
$r(\lambda)$,  we obtain a solution of (\req(STU))
for all
$\lambda$ with the new $\alpha_k,  0 \leq k \leq
\frac{d}{2}$, in the form :
\be
\alpha_k = P_k(\lambda) \ , \ \ {\rm deg}\  P_k
(\lambda) = [\frac{k}{2}] + [
\frac{d+2}{4}] - \frac{{\bf n}-1}{2} \ ; 
\ele(mP) 
in particular, ${\rm deg} \ 
P_\frac{d}{2}(\lambda)  = M$.  Now the constraint
(\req(QQdc)) becomes
$$
(\delta_{\frac{d}{2}+2}-\lambda)
P_{\frac{d}{2}}(\lambda) + 
2 u_{\frac{d}{2}+2}P_{\frac{d}{2}-1}(\lambda)
+ 2 \mu_{\frac{d}{2}+2}
P_{\frac{d}{2}-2}(\lambda)  = 0 , 
$$
by which one can show that the above relation gives
rise to
$(M+1)$ $\lambda$-values for a generic
$c$. 

In the case
(\ref{QQd-}),  (\req(QQdc))  becomes a trivial
relation. We consider the following eigenvalue
problem, similar to the one in (\req(STU)) by
changing
$S$ to
$S^{-}$:
\be
 \left( \begin{array}{cc}
S^{-}&T \\
0 &U
\end{array} 
\right)  
 \left( \begin{array}{c}
\widetilde{\psi} \\
\psi
\end{array} 
\right) = \vec{0} \ , \ \ \ \ \ \
\psi:= \left( \begin{array}{c}
\alpha_{{\bf n}-1} \\
\vdots\\
\vdots\\
\alpha_1 \\
\alpha_0
\end{array} 
\right) \ , \ \ \
\widetilde{\psi}:= 
\left( \begin{array}{c}
\alpha_{\frac{d}{2}} \\
\vdots \\
\vdots \\
\alpha_{{\bf n}+1} \\
\alpha_{\bf n} 
\end{array} 
\right) \ ,
\ele(S-TU)
where the matrix $S^{-}$ differs from $S$ only on
the $(1,2)$-th entry by changing 
$\nu_{\frac{d}{2}+1}$ to $-\nu_{\frac{d}{2}+1}$,
i.e.,
$$
S^{-}= \left( \begin{array}{ccccccc}
v_{\frac{d}{2}+1} &-\nu_{\frac{d}{2}+1}+
\delta_{\frac{d}{2}+1}-\lambda
&u_{\frac{d}{2}+1}&\mu_{\frac{d}{2}+1}&0 &
\cdots &0
\\
\nu_{\frac{d}{2}} &v_{\frac{d}{2}}&
\delta_{\frac{d}{2}}-\lambda
&u_{\frac{d}{2}}&\mu_{\frac{d}{2}}&0 & \vdots
\\
0& \ddots & & & & &
\\
\vdots &0 &\ddots & & & & \vdots 
\\
\vdots &\cdots &\cdots & & & &  
\\
\vdots &\cdots &0&\nu_{{\bf n}+3}&v_{{\bf n}+3}&
\delta_{{\bf n}+3}-\lambda &u_{{\bf n}+3}
\\
0&\cdots& \cdots &0&\nu_{{\bf n}+2}&v_{{\bf n}+2}&
\delta_{{\bf n}+2}-\lambda
 \\ 
0& \cdots & \cdots & & 0 &\nu_{{\bf n}+1}&v_{{\bf
n}+1} 
\end{array} 
\right) \ .
$$
Then the problem (\req(S-TU)) with the condition
$\alpha_{\frac{d}{2}}=0$ is equivalent to the
problem in the case (\ref{QQd-}). As in the
discussion of the eigenvalue problem
(\req(STU)), there is a solution  of
(\req(STU)), $\alpha_j$s, with the form $\alpha_k=
P_k^-(\lambda)$ satisfying the property 
(\req(mP)). Then
$P_{\frac{d}{2}}^- $ is a degree $M$
polynomial of $\lambda$, and its zeros, $
\alpha_{\frac{d}{2}} = P_{\frac{d}{2}}^-
(\lambda) = 0$,
give rise to $M$ $\lambda$-values  for the
case  (\ref{QQd-}). The  conclusion now follows
from Theorem \ref{thm:syQ}.

(III) $m=M$. We have 
$l=0,1$, and $d= N-1$.  As the values of
  $\nu_1, v_1, u_{N+2},
\mu_{N+2}$ in (\req(coef5)) are all zeros in this
case, the relations of
$j=1, d+3$ in (\req(eqDL4)) are  redundant. Hence
the system (\req(eqDL4)) is equivalent to the
eigenvalue problem (\req(Mform)) with 
$d=N-1$. In the case (\ref{QQd}),  the collection
of
$\alpha_k, 0\leq k\leq \frac{d}{2}$, among the
coefficients of
$Q(x)$  is the solution of the following
eigenvalue problem :
$$
\left\{ \left( \begin{array}{ccccccc}
\delta_{M+2} & 2 u_{M+2}
&2 \mu_{M+2}&0&\cdots  & 0 & 0 \\
v_{M+1}&
\delta_{M+1}+ \nu_{M+1}
&u_{M+1}&\mu_{M+1}&\ddots   &
0 &0 \\
\nu_M&v_M& \delta_M
&u_M&\mu_M&\ddots   &0 \\ 0&
\ddots&
\ddots&\ddots &\ddots  & \ddots &\vdots \\
\vdots& \ddots & \ddots & \ddots & \ddots & 
\ddots &0\\
\vdots& \ddots &\nu_4 &v_4 & \delta_4 
&u_4&\mu_4\\
\vdots& \ddots  & \ddots &\nu_3 &v_3 & \delta_3 
&u_3\\
0&  \cdots & & 0 &\nu_2& v_2 & \delta_2
\end{array} 
\right) - \lambda \right\}
 \left( \begin{array}{c}
\alpha_M\\
\alpha_{M-1} \\
\vdots \\
\vdots \\
\vdots \\
\vdots\\
\alpha_0
\end{array} 
\right) = \vec{0} \ .
$$
Note that the coefficients in the first and second
rows have some extra terms, comparing to the
rest of entries.  There are
$(M+1)$ $\lambda$-eigenvalues
 for the above relation, which gives rise to the
solutions for  the case (\ref{QQd}).  The
rest
$M$ $\lambda$-values for (\req(Mform)) 
are those for the case
(\ref{QQd-}). Then the conclusion follows
 from Theorem \ref{thm:syQ}.
$\Box$ \par \vspace{0.2in} \noindent
{\bf Remark} By Theorem
\ref{thm:SMTQ}, an eigen-polynomial 
$Q(x)$ of the  symmetric
$T$-$Q$ polynomial equation
$(\ref{BEml})$ are in sectors $(m,l)=(m, 0), (m,
N-2m)$, and it  has the form,
$$
Q(x) = \prod_{j=1}^{2N-2-2m} (x-\frac{1}{z_j}) \ ,
\
\ z_j \neq 0  \ .
$$
The Bethe ansatz equation (\ref{rtBZ}) of  $z_k$s
 becomes
\begin{eqnarray*}
 (\frac{z_j^2 + 2 {\rm i}c
q^{\frac{1}{2}} z_j -  q}{ q z_j^2 - 2  {\rm i} c
q^{\frac{1}{2}}z_j
 - 1})^2 = \prod_{n \neq j, n=1}^{2N-2-2m} 
\frac{z_n-q z_j}{qz_n - z_j} \ , \ \ \ \ \ 1 \leq
j \leq 2N-2-2m \ .
\end{eqnarray*}
By the description of $Q(x)$ in
Theorem 
\ref{thm:SMTQ}, one has the
reciprocal constraint on $ z_j$'s, i.e., 
$\{ z_j \}_{j=1}^{2N-2-2m} = \{ z_j^{-1}
\}_{j=1}^{2N-2-2m} $ (counting the
multiplicity).
$\Box$ \par \vspace{0.2in} \noindent

\section{The General Spectral Curve for  Discrete
Quantum Pendulum and Discrete Sine-Gordon Model} 
In this section, we are going to
explore the geometrical structure of
the spectral curve ${\cal C}_{\vec{h}}$
(\ref{eq:Cvh}) for the 
discrete quantum pendulum and 
SG model. By (\ref{condPSC}) and
(\req(DQPSG)), the parameter 
 $\vec{h}$ for the curve ${\cal C}_{\vec{h}}$ has
the following constraints,
\bea(lll)
 a_j =
q^{-1}d_j ^{-1} ,
& b_j = -k^{-\epsilon_j} c_j^{-1} , & 
(\epsilon_j := (-1)^j ) , \ 0 \leq j \leq 3 ,  \\
 d_0d_1d_2d_3=1,
& c_1^N c_3^N = k^{2N}c_0^Nc_2^N  \ . &
\elea(paSG)
By the discussion in Sect. 8 of \cite{LR}, the
value $\xi_j^N$s of the curve ${\cal C}_{\vec{h}}$
are determined by $x^N, \xi_0^N$, which we denote
by $y : = x^N, \eta :=\xi_0^N$. The variables $(y,
\eta)$ satisfies the following equation of the
curve ${\cal B}_{\vec{h}}$,
$$
C_{\vec{h}}(y) \eta^2 +
(A_{\vec{h}}(y)-  D_{\vec{h}}(y))\eta 
- B_{\vec{h}}(y) = 0 \ ,
$$
where $A_{\vec{h}}, B_{\vec{h}}, 
C_{\vec{h}}, D_{\vec{h}}$ are polynomials of
$y$ given by the relation, 
$$
\left( \begin{array}{cc}
-A_{\vec{h}}(y)  &  B_{\vec{h}}(y)  \\
C_{\vec{h}}(y)  & - D_{\vec{h}}(y)    
\end{array} \right) :
= \prod_{j=0}^3  \left( \begin{array}{cc}
- d_j^{-N}  &  -yc_j^{-N}k^{-\epsilon_jN}  \\
 y c^N_j  &-d^N_j    
\end{array} \right) \ .
$$
In fact, by computation,  one has the following
expressions of these polynomials,
$$
\begin{array}{ll}
A_{\vec{h}}(y) &=  -( \delta^{-1}-y^2 \gamma^{-1})
(\delta -y^2
\gamma)+y^2k^{-N}c_0^{-N}d_0^{-N}c_2^Nd_2^N(
\delta + k^{N}\gamma )^2 ; \\
D_{\vec{h}}(y) & = -(\delta^{-1}-y^2 \gamma^{-1})
(\delta -y^2
\gamma) +
y^2k^Nc_0^Nd_0^Nc_2^{-N}d_2^{-N}(k^{-N}
\delta^{-1} +  \gamma^{-1})^2 ; 
\\
B_{\vec{h}}(y)  &= y(
\delta^{-1}-y^2
\gamma^{-1})\{ k^{-N}c_0^{-N}d_0^{-N}  (
\delta + k^N \gamma )+
c_2^{-N}d_2^{-N}
(k^{-N}
\delta^{-1} +  \gamma^{-1}) \} ;  \\
C_{\vec{h}}(y)  & = 
-y(\delta -y^2 \gamma) \{ k^N c_0^Nd_0^N
( k^{-N}
\delta^{-1} +  \gamma^{-1}) + c_2^Nd_2^N
( 
\delta + k^N\gamma ) \} \ ,
\end{array}
$$ 
where $\delta, \gamma$ are defined
by  
$$
\delta:= d_0^Nd_1^N = d_2^{-N}d_3^{-N} \ , \ \
\gamma:=
\frac{c_0^Nk^N}{c_1^N}= \frac{c_3^N}{k^Nc_2^N} \ .
$$
Eliminating  the $y$-factor in
the equation of ${\cal B}_{\vec{h}}$, we
obtain an irreducible curve, which will still be
denoted by ${\cal
B}_{\vec{h}}$ for the convenience of notations,
but now with the equation:
\bea(l)
{\cal B}_{\vec{h}} : \  {\sf a} (y^2 \gamma-\delta
)
\eta^2 + {\sf b} y \eta + {\sf c}  (y^2
\gamma^{-1}-
\delta^{-1}) = 0 \ , 
\elea(Bvh)
where ${\sf a}, {\sf b}, {\sf c}$ are the
parameters defined by
$$
\begin{array}{ll}
{\sf a} :=  & k^N c_0^Nd_0^N
( k^{-N}
\delta^{-1} +  \gamma^{-1}) + c_2^Nd_2^N
( 
\delta + k^N\gamma ) \ , \\
{\sf b} := & k^{-N}c_0^{-N}d_0^{-N}c_2^Nd_2^N(
\delta + k^{N}\gamma )^2 -
k^Nc_0^Nd_0^Nc_2^{-N}d_2^{-N}(k^{-N}
\delta^{-1} +  \gamma^{-1})^2 \ ,   \\
{\sf c} := &k^{-N}c_0^{-N}d_0^{-N}  (
\delta + k^N \gamma )+
c_2^{-N}d_2^{-N}
(k^{-N}
\delta^{-1} +  \gamma^{-1}) .
\end{array}
$$
The
curves ${\cal B}_{\vec{h}}$ form a family of
elliptic curves, depending on the four parameters,
$\delta, \gamma, k^Nc_0^Nd_0^N, c_2^Nd_2^N$. 
And
${\cal C}_{\vec{h}}$ is a $\ZZ_N^5$-cover
(branched) over
${\cal B}_{\vec{h}}$, with the covering
transformation group containing  $\tau_\pm$  in
(\ref{DelTau}). For a generic $\vec{h}$, 
${\cal C}_{\vec{h}}$ is a high-genus curve; indeed
the genus is equal to
$2N^3(N-1)(N+2)+1$. 

Now we are going to make a qualitative analysis
on solutions of the Baxter's $T$-$Q$ relation
(\req(BetheT*)) related to the  discrete
quantum pendulum (\req(DQP)) and  SG model
(\req(SG)). All the linear transformations
appeared in the expressions of (\req(TTsp)) are
operators of the vector space
$\stackrel{4}{\otimes}\CZ^{N*}$. As $D, C$ and
$U_j$s in the expression of $T^\ast_{\vec{h}}(x)$
are commuting operators, by
$(\frac{k^2c_0c_2}{c_1c_3}C)^N=1$, the eigenvalue
of $T^\ast_{\vec{h}}(x)$ are still expressed in
the form of (\req(lambda)) with  $\lambda$ 
depending on $k$; while for the discrete quantum
pendulum and SG model, it becomes
(\req(lmbasy)). By the expressions of $D$ and $
C$, it is not hard to see that the common
eigen-subspaces of
$\stackrel{4}{\otimes}\CZ^{N*}$ for the commuting
operators $D^{\frac{1}{2}},
\frac{k^2c_0c_2}{c_1c_3}C$  all have the dimension
$N^2$. The eigenspace decomposition of
$\stackrel{4}{\bigotimes}\CZ^{N*}$ is denoted  by
$$
\stackrel{4}{\bigotimes}\CZ^{N*} = \bigoplus_{n,
n' \in \ZZ_N} {\bf E}_{n, n'} \ , \ \ \ \ {\bf
E}_{n, n'}
\simeq \CZ^{N^2} \ , 
$$
where $D^{\frac{1}{2}},
\frac{c_1c_3}{k^2c_0c_2}C^{-1}$ act on ${\bf
E}_{n, n'}$ by the multiplication of 
$q^n, q^{n'}$ respectively.  
By the relations of $U_j$'s and $C, D$
in (\req(UV)), each ${\bf E}_{n, n'}$ is stable
under
$U_j$s. The operators $U_j$s on  ${\bf
E}_{n, n'}$ are determined only by those of
$U_1, U_2$, which form the Weyl algebra: $
 U_2U_1 = \omega U_1U_2$ , $ U_1^N = U_2^N=
1 $. 
This implies the operator $T_2^\ast$ 
in (\req(TTsp)) is determined by the
representation ${\bf E}_{n, n'}$ of the Weyl 
algebra  on each  sector, labelled by the
eigenvalues  of
$T_0^\ast, T_4^\ast$ corresponding
to $(n, n')$. As the irreducible representation
of Weyl algebra  is unique, given
by the standard one on $\CZ^N$,
${\bf E}_{n, n'}$ is isomorphic to the sum of
$N$-copies of the standard representation as
Weyl algebra modules. In particular, the
eigenvalues of $-T_2^\ast$ on the vector
space ${\bf E}_{n, n'}$ are induced from
the standard representation of the
Weyl algebra ; each eigenvalue
gives rise to $N$ eigenvectors in
${\bf E}_{n, n'}$. 
By (\req(Bv)) and (\req(paSG)), 
the Baxter vacuum
state $|p
\rangle \in \ 
\stackrel{4}{\otimes} \CZ^N $ for $p \in {\cal
C}_{\vec{h}}$ is now defined by  $|p\rangle =
|p_0\rangle \otimes |p_1\rangle \otimes
|p_2 \rangle  \otimes |p_3
\rangle$, where the vector $| p_j\rangle$
in $\CZ^N$ is given by the conditions:
$$
\langle 0| p_j \rangle = 1 \ , \ \ \ \
\frac{\langle m|p_j \rangle}{\langle
m-1|p_j \rangle} = 
\frac{\xi_{j+1}k^{\epsilon_j} c_j q^{2m-1}  + x
d_j
 }{ - \xi_j (  \xi_{j+1} x c_j
q^{2m} - d_j) k^{\epsilon_j} c_j d_j }  \ .
$$
For a generic $\vec{h}$, the evaluation of 
vectors of $\stackrel{4}{\otimes} \CZ^{* N}$ on
the Baxter vacuum state, $* \mapsto \langle *|p
\rangle$, induces an isomorphism between
$\stackrel{4}{\otimes} \CZ^{* N}$ and a
$N^4$-dimensional subspace of rational functions
of ${\cal C}_{\vec{h}}$. Through this
isomorphism,
${\bf E}_{n, n'}$ gives rise a
$N^2$-dimensional functional space of
${\cal C}_{\vec{h}}$,
denoted by
$\epsilon({\bf E}_{n, n'})$, with a
Weyl-algebra-module structure induced from ${\bf
E}_{n, n'}$. In the Baxter's $T$-$Q$ equation
(\req(BetheT*)) on ${\cal C}_{\vec{h}}$ with
$\Lambda^*(x)=
\Lambda_{m, l}(x)$ in (\req(lambda)), the
function $Q(p)$ is the
$T^\ast_{\vec{h}}(x)$-eigenfunction  
 in $\epsilon({\bf E}_{n,
n'})$ for $ (n, n')=(m, l),  (N-m, N-l)$, with
the multiplicity $N$. For
the discrete quantum pendulum and SG model, by
(\req(DQP)) and (\req(SG)),  the
$\Lambda^*(x)$ in Baxter's $T$-$Q$
relation (\req(BetheT*)) is the reciprocal
polynomial (\req(lmbasy)). To determine
the eigen-functions $Q(p)$ would require the
understanding of its zeros and poles by using the
expression of the Baxter vacuum state, which has
been a difficult task at this moment. The possible
role of elliptic function theory of
${\cal B}_{\vec{h}}$ in the solutions of  
Baxter's $T$-$Q$ relation on ${\cal C}_{\vec{h}}$,
and some further understanding on the eigenvalue
problem of the models (\req(DQP)) (\req(SG))
along this line, would be the core of our future
work in this aspect.

\section{Conclusions and Perspectives}
We have studied the discrete quantum pendulum
and discrete sine-Gordon model in the framework
of quantum inverse scattering method. The
diagonalization problem is governed
by  the Baxter's $T$-$Q$ relation, which arises
from the Baxter vacuum state on the
spectral curve through a general scheme
by using the transfer matrix for a
fixed finite size $L$. We have
demonstrated the role of algebraic geometry in
the qualitative study of the Baxter's $T$-$Q$
relation for
$L$=3 in
\cite{LR}, and $L$=4 in this article. 
In both cases, they have shown an intimate
relationship with integrable Hamiltonian
spin-chains of physical interest. In this
approach, one relies on the spectral curves,
depending on the parameters encoded in the
corresponding Hamiltonion expression. For 
generic parameters, the spectral curve, where the
Baxter's
$T$-$Q$ relation is formulated, is  a high-genus
Riemann surface, as demonstrated in Sect. 6.
However for both $L=3$ and $4$, the spectral
curves possess a common feature that they form
certain branched covers over elliptic curves. One
might expect to employ the elliptic function
theory to the solutions of Baxter's
$T$-$Q$ relation so that the algebraic geometry
study could enrich our understanding of the 
corresponding Hamiltonian spectrum problem.
Although this thinking is merely a speculation at
present, a possible program along this line
could be  a challenging one,  on which we hope to
make progress in future.

When the spectral curve degenerates
into rational curves, where the geometry plays
little role, 
we derive the polynomial formulation of 
Baxter's
$T$-$Q$ relation for a system of
a finite size $L$ in Sect. 3. We
apply these results  to the case 
$L$=4 in Sect. 4 for the
setting of  discrete quantum
pendulum and sine-Gordon model. In these cases, 
an extra symmetry has naturally been imposed on
the Baxter's
$T$-$Q$ polynomial equation; indeed it is
governed by the
reciprocal property of the equation. We present a
detailed and  rigorous mathematical derivation of 
the solutions in Theorem 
\ref{thm:SMTQ}. Surprisingly
the conclusion on these
polynomial  solutions has been much in tune with
the one for
$L=3$  on the study of
Hofstadter-type model  (see Theorem 3 in
\cite{LR}). Furthermore, the exact connection of
the Baxter's
$T$-$Q$ polynomial equation with the Bethe ansatz
technique in literature has been clarified
in all these cases.
The results obtained 
in this paper signal some further mathematical
feature of the  Baxter's
$T$-$Q$ polynomial equation, namely, the novel
connection with certain
$q$-Sturm-Liouville problem at roots of unity
$q^N=1$. 
The facts discovered in this work could
be served to demonstrate that a systematic
mathematical theory  embodied in  the
Baxter's $T$-$Q$ polynomial equation (or
algebraic Bethe Ansatz) would emerge  in
the study of $q$-difference
operators.  Accordingly, the relationship along
this line is now under our consideration.

\section*{ Acknowledgments} 
We would like to thank R. Seiler for informing
us his joint work with J. Kellendonk and N. Kutz
\cite{KKS}. The second author would acknowledge
informative communications with G. von Gehlen on
Baxter's $T$-$Q$ relation during his visit of
Physikalishe Institut der Universit\"{a}t Bonn,
Germany, in June, 2001.  S. S. Roan is supported
in part by the National Science Council of Taiwan
under grant No. 89-2115-M-001-037.

\section*{ Appendix: Discrete Quantum Sine-Gordon
Hamiltonian}
For the consistency of our notion of 
discrete quantum sine-Gordon Hamiltonian with the
ones used in other literature,  we  make an 
identification of the discrete  sine-Gordon
integral in 
\cite{BKP} with the $T_j^*$s of (\req(SG)) in
this paper.  In  Sect. 5 of \cite{BKP}, the
discrete quantum sine-Gordon Hamiltonian arises
from the following commuting 
operators\footnote{Here we use the sans serif
type style, instead of the italic type style in
\cite{BKP}, for operators appeared in the
right hand side of the expressions, for the
purpose of less confusion on notations used in
this paper.},
$$
\begin{array}{ll}
A^{(0)} = & {\sf U}_2^{-1} {\sf U}_1^{-1} , \\
A^{(1)} = & {\sf U}_2{\sf Z}_2{\sf U}_1^{-1} +
{\sf U}_1{\sf Z}_1{\sf U}_2^{-1}+ {\sf V}_2 {\sf
h}_2 {\sf h}_1^* {\sf V}_1^{-1} \ ,
\\ 
A^{(2)} =&  {\sf U}_2 {\sf Z}_2 {\sf U}_1 {\sf
Z}_1 , 
\end{array}
$$
where the lower index $j=1,2$ indicates the site
of operators in the same algebra generated by
${\sf U}, {\sf V}, {\sf Z},$ subject to the
relation ${\sf U}{\sf V}= {\sf q}^{\frac{-1}{2}}
{\sf V}{\sf U}$ with ${\sf Z}$ the central
element, and ${\sf h}$ is defined by
 ${\sf h} = k^{\frac{-1}{2}}+ k^{\frac{1}{2}}
{\sf q}^{\frac{-1}{2}} {\sf U}^2 {\sf Z}$.
The sine-Gordon $(SG)$ integral is defined by the
operator
$$
\widetilde{H} = A^{(1)} + A^{(1) *} \ .
$$
One can show that 
$$
{\sf V}_2 {\sf
h}_2 {\sf h}_1^* {\sf V}_1^{-1} = {\sf
q}^{\frac{-1}{2}}{\sf V}_2 {\sf U}_2^2 {\sf Z}_2
{\sf V}_1^{-1} + {\sf
q}^{\frac{1}{2}}  {\sf V}_2 {\sf Z}_1^{-1} {\sf
U}_1^{-2}{\sf V}_1^{-1} + k {\sf V}_2{\sf
U}_2^2
{\sf Z}_2 {\sf Z}_1^{-1} {\sf U}_1^{-2} {\sf
V}_1^{-1}  + k^{-1} {\sf V}_2 {\sf V}_1^{-1} \ .
$$
For the convenience in expressing 
$\widetilde{H}$, we denote 
$$
\begin{array}{ll}
{\sf W}_1 : = {\sf
q}^{\frac{1}{2}}  {\sf V}_2 {\sf Z}_1^{-1} {\sf
U}_1^{-2}{\sf V}_1^{-1} , & {\sf W}_2 : = {\sf
U}_1{\sf Z}_1{\sf U}_2^{-1} \ , 
 \\ {\sf W}_3 : = {\sf
q}^{\frac{-1}{2}}{\sf V}_2 {\sf U}_2^2 {\sf Z}_2
{\sf V}_1^{-1} \ , & {\sf W}_4 : = {\sf U}_2{\sf
Z}_2{\sf U}_1^{-1} \ .
\end{array}
$$ 
Note that 
\be
{\sf W}_2{\sf W}_4 = A^{(0)}A^{(2)} \ .
\ele(WW)
We have 
$$
\begin{array}{ll}
k {\sf V}_2{\sf
U}_2^2
{\sf Z}_2 {\sf Z}_1^{-1} {\sf U}_1^{-2} {\sf
V}_1^{-1} =  k {\sf q}^{\frac{-1}{2}}  A^{(0)}
{\sf W}_3 {\sf W}_2^{-1}  , & 
 k^{-1} {\sf V}_2
{\sf V}_1^{-1} = k^{-1} {\sf
q}^{\frac{-1}{2}}A^{(2)} {\sf W}_4^{-1}{\sf W}_1 ,
\end{array}
$$
hence
$$
A^{(1)}= {\sf W}_1 + {\sf W}_2 + {\sf W}_3 +{\sf
W}_4 + k {\sf q}^{\frac{-1}{2}}  A^{(0)}
{\sf W}_3 {\sf W}_2^{-1} + k^{-1} {\sf
q}^{\frac{-1}{2}}A^{(2)} {\sf W}_4^{-1}{\sf W}_1
\ .
$$
With ${\sf q}^{\frac{1}{2}}=q$, we identify the
operators appeared in the above $A^{(j)}$s with
those in 
$T^*_{2j}$s under the condition
$d_0d_1d_2d_3=1$,
$$
\begin{array}{llllll}
A^{(0)} & \leftrightarrow &
D^{\frac{1}{2}} \  ;&  A^{(2)} & \leftrightarrow
&  \ 
\frac{c_1c_3}{k^2c_0c_2} D^{\frac{1}{2}} C^{-1}
;
\\ {\sf W}_1 & \leftrightarrow &
\frac{kc_0d_2d_3}{c_1} D^{\frac{-1}{2}} U_1 ;
&
{\sf W}_2 & \leftrightarrow &
\frac{kc_2}{c_1d_0d_3} D^{\frac{1}{2}}
U_2^{-1} ; \\ {\sf W}_3 & \leftrightarrow &
\frac{kc_2d_0d_1}{c_3} D^{\frac{-1}{2}} U_3 ;
&
{\sf W}_4 & \leftrightarrow &
\frac{c_3d_1d_2}{kc_0} D^{\frac{-1}{2}} U_4 , 
\end{array}
$$
and then impose further constraints,
\be
{\sf W}_4^{-1} \ \leftrightarrow \
\frac{c_1d_0d_3}{kc_2} D^{\frac{-1}{2}} U_2 \ , \
\ 
\ 
A^{(2)} \ \leftrightarrow \ D^{\frac{1}{2}} \ .
\ele(extSG)
By the equalities $V_1=U_3U_2,
V_4=U_2U_1$ in (\req(UV)), the $SG$-integral
$\widetilde{H}$ becomes $-T^*_2$ in
(\req(SG)). Then (\req(extSG)) gives
rise to the identification, 
$$
{\sf W}_2 = {\sf W}_4 \ , \ \ A^{(0)}= A^{(2)} \
, 
$$
or equivalently, 
$$
\frac{kc_2}{c_1d_0d_3} D^{\frac{1}{2}} U_2^{-1} =
\frac{c_3d_1d_2}{kc_0} D^{\frac{-1}{2}} U_4 \ , \
\ \ 
\ \ \frac{c_1c_3}{k^2c_0c_2} = C \ .
$$
Note that the above relations are consistent with
the relation
$U_2U_4=C^{-1}D$ in (\req(UV)).
Hence we obtain the relations (\req(DQPSG))
(\req(SG)).


\begin{thebibliography}{99}
\bibitem{A} M.Ya. Azbel, Energy spectrum of a
conduction electron in a magnetic field, Sov. Phys. JETP
19 (1964) 634-645.


\bibitem{B} R. J. Baxter, Partition function of
the eight-vertex lattice model, Ann. Phys. 70
(1972) 193-228; eight-vertex model in lattice
statistics and one-dimensional anisotropic
Heisenberg chain, I, II, III, Ann. Phys. 76
(1973) 1-24, 25-47, 48-71.

\bibitem{Bax} R. J. Baxter, Exactly solved models
in statistical mechanics, Academic Press (1982).

\bibitem{BBP} R. J. Baxter, V.V. Bazhanov and
J.H.H. Perk,  Functional relations for transfer
matrices of the chiral Potts model, Int. J. Mod.
Phys. B 4 (1990) 803-870.


\bibitem{BazS} V.V. Bazhanov and Yu.G. Stroganov, Chiral
Potts model as a descendant of the six-vertex model, J.
Stat. Phys. 59 (1990) 799-817.

\bibitem{BKP} A. Bobenko, N. Kutz and U. Pinkall, The
discrete quantum pendulum, Phys. Letts. A 177 (1993)
399-404.


\bibitem{F} L. D. Faddeev, How algebraic Bethe
Ansatz works for integrable models, eds. A.
Connes, K. Gawedzki and J. Zinn-Justin, {\it
Quantum symmetries/ Symmeties quantiques},
Proceedings of the Les Houches summer school,
Session LXIV, Les Houches, France, August 1-
September 8, 1995, North-Holland (1998) 149-219;

L. A. Takhtadzhan and L. D.
Faddeev , The quantum method of the inverse
problem and the Heisenberg XYZ model, Russ.
Math. Surveys 34 ( 1979 ) 11-68.


\bibitem{FK} L. D. Faddeev and R. M. Kashaev, 
Generalized
Bethe Ansatz equations  for Hofstadter problem, 
Comm.
Math. Phys. 155 (1995) 181-191, hep-th/9312133.

\bibitem{H} D. R. Hofstadter, Energy levels and wave
functions of Bloch electrons in rational and
irrational magnetic fields, Phys. Rev. B 14 (1976)
2239-2249.


\bibitem{KKS} J. Kellendonk, N.  Kutz and R. Seiler, 
Spectra of quantum integrals, in {\it Discrete
Integrable Geometry and Physics }, eds. A I
Bobenko and R. Seiler, Oxford Lectures Series in
Mathematics and Applications, No 16, Oxford
University Press Inc., New York, 1999, pp.
247-297.

\bibitem{KS} P. P. Kulish and E. K. Sklyanin,
Quantum spectral transform method. Recent
developments, eds. J. Hietarinta and C. Montonen,
Lecture Notes in Physics 151 Springer (1982)
61-119.


\bibitem{IK} A. G. Izergin and V. E. Korepin,
Lattice versions of quantum field theory models
in two dimensions, Nucl. Phys. B 205 (1982) 401-
413;

V. O. Tarasov, Irreducible monodromy matrices for
$R$-matrix of the $XXZ$-model and lattice local
quantum Hamiltonians, Theor. Math. Phys. 63 (195)
175-196 (in Russian). 

\bibitem{LR} S. S. Lin and S. S. Roan, Algebraic
geometry approach  to the Bethe equation for
Hofstadter type models, J.  Phys. A: Math. Gen. 35
(2002) 5907-5933, cond-mat/9912473;  Algebraic
Geometry and Hofstadter Type Model,  Intern. J. 
Mod.  Phys. B,  16 (2002) 2097-2106,  
math-ph/0206014 .






\end{thebibliography}
\end{document}